\documentclass[aps,prd,reprint,superscriptaddress,nofootinbib]{revtex4-2}
\usepackage{amssymb,amsmath,bm,natbib}
\usepackage[dvipsnames]{xcolor}
\usepackage[colorlinks = true,
            linkcolor = Maroon,
            urlcolor  = Maroon,
            citecolor = Maroon]{hyperref}
\usepackage{microtype}
\usepackage{graphicx}
\usepackage{xspace}
\usepackage{bm}
\usepackage{bbm}
\usepackage{tikz}
\usetikzlibrary{decorations.markings}
\usetikzlibrary{decorations.pathmorphing}
\tikzset{snake it/.style={decorate, decoration=snake}}
\usetikzlibrary{matrix,calc}

\renewcommand{\arraystretch}{1.5}

\begin{document}

\title{Optimal anti-ferromagnets for light dark matter detection}

\author{Angelo~Esposito}
\email{angelo.esposito@uniroma1.it}
\affiliation{Dipartimento di Fisica, Sapienza Universit\`a di Roma, Piazzale Aldo Moro 2, I-00185 Rome, Italy}
\affiliation{INFN Sezione di Roma, Piazzale Aldo Moro 2, I-00185 Rome, Italy}
\affiliation{School of Natural Sciences, Institute for Advanced Study, Princeton, NJ 08540, USA}

\author{Shashin~Pavaskar}
\email{spavaska@andrew.cmu.edu}
\affiliation{Department of Physics, Carnegie Mellon University, Pittsburgh, PA 15213, USA}

\date{\today}

\begin{abstract}
\noindent We propose anti-ferromagnets as optimal targets to hunt for sub-MeV dark matter with spin-dependent interactions. These materials allow for multi-magnon emission even for very small momentum transfers, and are therefore sensitive to dark matter particles as light as the keV. We use an effective theory to compute the event rates in a simple way. Among the materials studied here, we identify nickel oxide (a well-assessed anti-ferromagnet) as an ideal candidate target. Indeed, the propagation speed of its gapless magnons is very close to the typical dark matter velocity, allowing the absorption of all its kinetic energy, even through the emission of just a single magnon.
\end{abstract}

\maketitle

%%%%%%%%%%%%%%%%%%%%%%%%%%%%%%%%%%%%%%%%%%%%%%%%%%%%%%%%%%%

\section{Introduction}

There is today overwhelming evidence that most of the matter in the Universe is dark. Despite that, the question about its nature arguably remains among the biggest ones in fundamental physics. In particular, the possible dark matter mass spans a range of several orders of magnitude. 
In light of stringent constraints on heavy WIMPs~\cite[e.g.,][]{SuperCDMS:2015eex,LUX:2016ggv,SuperCDMS:2018mne,XENON:2019gfn,CRESST:2019jnq,PandaX-II:2020oim}, recent years have witnessed an increasing interest in models for sub-GeV dark matter~\cite[e.g.,][]{Boehm:2003ha,Boehm:2003hm,Strassler:2006im,Hooper:2008im,Feng:2008ya,Falkowski:2011xh,Lin:2011gj,Hochberg:2014dra,DAgnolo:2015ujb,Kuflik:2015isi,Green:2017ybv,DAgnolo:2018wcn}, also motivating new detection ideas. In particular, dark matter candidates in the keV to GeV range, while still heavy enough to be considered as particles, cannot release appreciable energy via standard nuclear recoil. They therefore require detectors with low energy thresholds, such as semiconductors~\cite{Essig:2011nj,Graham:2012su,Essig:2015cda,Hochberg:2016sqx,Bloch:2016sjj,Knapen:2020aky,Liang:2022xbu,Berghaus:2022pbu}, superconductors~\cite{Hochberg:2015pha,Hochberg:2016ajh,Hochberg:2019cyy,Griffin:2020lgd,Hochberg:2021yud}, Dirac materials~\cite{Hochberg:2017wce,Coskuner:2019odd,Geilhufe:2019ndy}, lower dimensional materials~\cite{Capparelli:2014lua,Hochberg:2016ntt,Cavoto:2016lqo,Cavoto:2017otc}, and so on (see also~\cite{Arvanitaki:2017nhi,Bunting:2017net,Chen:2020jia}).

Among these, the proposals based on superfluid ${^4}\text{He}$~\cite{Guo:2013dt,Schutz:2016tid,Knapen:2016cue,Hertel:2018aal,Acanfora:2019con,Caputo:2019cyg,Caputo:2019xum,Baym:2020uos,Caputo:2020sys,Matchev:2021fuw,You:2022pyn,vonKrosigk:2022vnf,Seidel:2022ofd} and solid crystals~\cite{Knapen:2017ekk,Griffin:2018bjn,Campbell-Deem:2019hdx,Cox:2019cod,Campbell-Deem:2022fqm} aim at detecting the collective excitations (phonons) produced by the spin-\emph{independent} interaction of dark matter with the nuclei in the material---for an overview see~\cite{Griffin:2019mvc,Trickle:2019nya,Kahn:2021ttr}. These collective modes have typical energies below $\mathcal{O}$(100~meV), and are therefore sensitive to particles as light as $m_\chi\sim\mathcal{O}(\text{keV})$. Different proposals for the detection of single phonons have been recently put forth~\cite{Maris:2017xvi,Osterman:2020xkb,Lyon:2022sza,Das:2022srn}.

The targets above are, however, not the most suitable ones to probe possible scenarios where spin-\emph{dependent} interactions of dark matter with the Standard Model are dominant over the spin-independent ones. In this regard, it has been proposed to use ferromagnets~\cite{Trickle:2019ovy,Mitridate:2020kly,Chigusa:2020gfs}, i.e. materials that exhibit a non-zero macroscopic magnetization in their ground state.\footnote{The materials presented in~\cite{Trickle:2019ovy,Mitridate:2020kly} are actually insulating ferrimagnets. This makes no difference in our discussion~\cite{Pavaskar:2021pfo}. We refer to ferromagnets, which are conceptually simpler.} The dark matter can interact with the individual spins in the target, exciting their local precession: a propagating collective mode called magnon. The proposals to detect single magnons involve either calorimetric readout~\cite{Trickle:2019ovy}, using TES or MKID, or quantum sensors, which instead couple the magnon mode to a superconducting qubit~\cite{doi:10.1126/science.aaz9236,lachance2019hybrid,doi:10.1126/sciadv.1603150}. A generic ferromagnet features several magnon types (branches). However, for sufficiently light dark matter ($m_\chi \lesssim 10$~MeV, for the typical material~\cite{Trickle:2019ovy}), the momentum transfer becomes smaller than the inverse separation between the spins. In this regime the event rate is dominated by the emission of gapless magnons which, for ferromagnets, are characterized by a quadratic dispersion relation, $\omega(q) = q^2/(2m_\theta)$, with $m_\theta$ a mass scale set by the properties of the material under consideration. Moreover, as we argue below, conservation of total magnetization implies that, when only gapless magnons are allowed, no more than one can be produced in each event. Thus, for $m_\chi \lesssim 10$~MeV, the maximum energy that can be released to a ferromagnet is $\omega_{\rm max} = 4 T_\chi x/(1+x)^2$, with $T_\chi$ the dark matter kinetic energy and $x\equiv m_\theta/m_\chi$. Typically, $m_\theta \sim \mathcal{O}(\text{MeV})$  (e.g., $m_\theta\simeq 3.5$~MeV for $\text{Y}_3\text{Fe}_5\text{O}_{12}$~\cite{Trickle:2019ovy}, see also \cite{srivastava1987spin,PhysRevB.64.174402}), and a sub-MeV dark matter will not deposit all its energy to the target.

In this work, we show that, instead, \emph{anti}-ferromagnets are optimal materials to probe the spin-dependent interactions of light dark matter. Similarly to ferromagnets, they also exhibit magnetic order in the ground state, but the spins are anti-aligned, leading to a vanishing macroscopic magnetization. This leads to two crucial differences: (1) gapless magnons have a linear dispersion relation, $\omega(q) = v_\theta q$, and (2) the interaction with the dark matter can excite any number of them.
If only one magnon is emitted, the maximum energy that can be transferred to the anti-ferromagnet is $\omega_{1,\rm max} = 4T_\chi y(1-y)$, with $y\equiv v_\theta/v_\chi$. One of the anti-ferromagnets we consider here, nickel oxide, features magnons with a propagation speed surprisingly close to the typical dark matter velocity, which allows it to absorb most of the kinetic energy even through a single magnon mode. This is a well-known and well-studied material, which makes it a particularly ideal target. Moreover, the possibility of exciting several magnons in a single event relaxes the kinematic constraints above, allowing any anti-ferromagnet to absorb the totality of the dark matter kinetic energy, hence being sensitive to masses down to $m_\chi \sim \mathcal{O}(\text{keV})$.

In what follows, we describe anti-ferromagnets and their interaction with dark matter via an effective field theory (EFT)~\cite{Burgess:1998ku,Schutte-Engel:2021bqm,Pavaskar:2021pfo}. This elucidates the role played by conservation laws in allowing multi-magnon emission and allows the computation of the corresponding event rates in a simple way, bypassing the difficulties encountered with more traditional methods~\cite[e.g.,][]{squires1996introduction,lovesey1984theory,dyson1956general}.

\vspace{1em}

\noindent \emph{Conventions:} We work in natural units, $\hbar = c = 1$, and employ the indices $i,j,k=1,2,3$ for spatial coordinates and $a,b=1,2$ for the broken $SO(3)$ generators.

%%%%%%%%%%%%%%%%%%%%%%%%%%%%%%%%%%%%%%%%%%%%%%%%%%%%

\section{The EFT}

\subsection{Magnons alone}

\begin{figure}[t]
    \centering
    \includegraphics[width=0.2\textwidth]{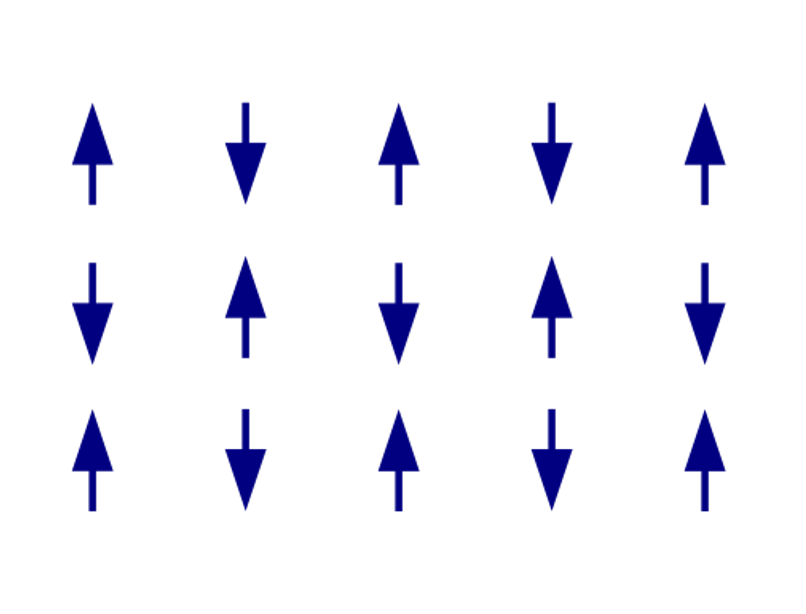}
    \caption{Schematic representation of the spins in the ground state of an anti-ferromagnet.}
    \label{fig:antiferromagnet}
\end{figure}

One can often picture an atom in a magnetic material as having a net spin coming from the angular momentum of the electrons localized around it. The Coulomb interaction between electrons pertaining to different atoms induces a coupling between different spins which, in turns, causes magnetic order in the ground state~\cite[e.g.,][]{ashcroft2011solid}. In an anti-ferromagnet these interactions are such that the spins are anti-aligned along one direction (Figure~\ref{fig:antiferromagnet}), which from now on we take as the $z$-axis. One can then define an order parameter, the so-called N\'eel vector, as $\bm{\mathcal{N}}\equiv \sum_i (-1)^i\bm{\mathcal{S}}_i$, where $\bm{\mathcal{S}}_i$ is the $i$-th spin and $(-1)^i$ is positive for those sites pointing `up' in the ordered phase, and negative for those pointing `down'. In the ground state the N\'eel vector acquires a non-zero expectation value, $\langle \bm{\mathcal{N}} \rangle \neq 0$~\cite[e.g.,][]{Burgess:1998ku}.

In the non-relativistic limit, a system of 3-dimensional spins enjoys an internal $SO(3)$ symmetry. The ground state described above breaks it spontaneously down to only the rotations around the $z$-axis, $SO(3)
\to SO(2)$, and the gapless magnons are nothing but the associated Goldstone bosons. As such, at sufficiently low energies, they are described by a universal EFT, very much analogous to the chiral Lagrangian in QCD.  A convenient way of parametrizing the magnons is as fluctuations of the order parameter around its equilibrium value, $\hat{\bm n} \equiv e^{i \theta^a S_a}\cdot \hat{\bm z}$, with $a=1,2$.
Here $\theta^a(x)$ is the magnon field and $S_a$ are the broken $SO(3)$ generators. 

The EFT Lagrangian is derived purely from symmetry considerations. First of all, one notes that under time reversal each spin changes sign, $\bm{\mathcal{S}}_i \to -\bm{\mathcal{S}}_i$. If combined with a translation by one lattice site, which swaps spin `up' with spin `down', this leaves the ground state unchanged. The effective Lagrangian for anti-ferromagnets must then be invariant under the joint action of these two symmetries. At large distances, translations by one lattice site do not affect the system, and the only requirement is time reversal: the Lagrangian must feature an even number of time derivatives~\cite{Burgess:1998ku}. 
Moreover, the underlying crystal lattice spontaneously breaks boosts. Assuming, for simplicity, that the material is homogeneous and isotropic at long distances, this implies that there must be explicit invariance under spatial translations and rotations, but that space and time derivatives can be treated separately.\footnote{We treat the underlying solid as a background which spontaneously breaks some spacetime symmetries. 
The corresponding Goldstone bosons, the phonons, realize these symmetries nonlinearly and can be included in the description if necessary \cite{Pavaskar:2021pfo}.} Since $|\hat{\bm n}| = 1$, the most general low-energy 
 Lagrangian for the gapless magnons\footnote{Crystalline anisotropy, which arises from spin-orbit coupling, can result in a small gap for magnons. This is analogous to the pion obtaining a mass due to the explicit breaking of chiral symmetry.} in an anti-ferromagnet is then~\cite{Burgess:1998ku,Pavaskar:2021pfo}, 
\begin{align} \label{eq:Ltheta}
    \begin{split}
        \mathcal{L}_\theta ={}& \frac{c_1}{2} {\big(\partial_t \hat{\bm n} \big)}^2 - \frac{c_2}{2} {\big(\partial_i \hat{\bm n} \big)}^2 \\
        ={}& \frac{c_1}{2} {\big(\dot \theta^a \big)}^2 - \frac{c_2}{2} {\big( \partial_i \theta^a \big)}^2 + \dots \,,
    \end{split}
\end{align}
where in the second equality we expanded in small fluctuations around equilibrium. The coefficients $c_{1,2}$ depend on the details of the anti-ferromagnet under consideration, and cannot be determined purely from symmetry. 

One recognizes Eq.~\eqref{eq:Ltheta} as the real representation of the Lagrangian of a complex scalar, corresponding to two magnons with linear dispersion relation, $\omega(q) = v_\theta q$, and propagation speed $v_\theta^2 = c_2/c_1$. The two magnons are completely analogous to relativistic particle and anti-particle, and they carry opposite charge under the unbroken $SO(2)$. As shown in~\cite{Burgess:1998ku,Pavaskar:2021pfo}, the action for a ferromagnet, instead, contains only one time derivative and it is analogous to that of a non-relativistic particle, which does not feature excitations with opposite charge. This is the reason why, when coupled to light dark matter, anti-ferromagnets allow for the emission of more than one magnon in each event, while ferromagnets do not. We discuss this more in Section~\ref{sec:DMmagnon}.

As far as our application is concerned, a central role is played by the spin density, which is the time-component of the Noether current associated to the original $SO(3)$ symmetry~\cite{Burgess:1998ku,Pavaskar:2021pfo}. This rotates the $\hat{\bm n}$ vector  (i.e., $\hat{n}_i \to R_{ij} \hat{n}_j$), and the current can be computed with standard procedures, giving the spin density:
\begin{align} \label{eq:s}
    \begin{split}
        s_i ={}& c_1 \big( \hat{\bm n} \times \partial_t \hat{\bm n} \big)_i = c_1 \Big( \delta_{ia} \dot \theta^a + \delta_{i3} \epsilon_{ab} \theta^a \dot \theta^b + \dots \Big) \,.
    \end{split}
\end{align}
From the equation above we also deduce that, while the ratio $c_2/c_1$ can be determined from the magnon speed, the coefficient $c_1$ can be found from an observable sensitive to the spin density of the anti-ferromagnet. One such quantity is the neutron scattering cross section, which we discuss in detail in Appendix~\ref{app:neutron}. 

Finally, our EFT breaks down at short wavelengths, when the dark matter is able to probe the microscopic details of the material. In other words, it loses validity for momenta larger than a certain strong coupling scale, $\Lambda_{\rm UV}$. The latter can be estimated, for example, as the momentum for which the dispersion relation sensibly deviates from linearity, which indicates that higher derivative terms in the Lagrangian~\eqref{eq:Ltheta} become relevant. In this work we consider three anti-ferromagnets: nickel oxide (NiO), manganese oxide (MnO) and chromium oxide (Cr$_2$O$_3$). In Table~\ref{tab:coeffs} we report their values of $v_\theta$, $c_1$, $\Lambda_{\rm UV}$, and of their density, $\rho_{\rm T}$.

\begin{table}[]
    \centering
    \begin{tabular}{c || c c c c}
         & $v_\theta$ & $c_1$ [MeV/\AA] & $\Lambda_{\rm UV}$ [keV] & $\rho_{\text{T}}$ [g/cm$^3$] \\ \hline\hline
         NiO \!\cite{hutchings_inelastic_1971} & \hspace{0.05em} $1.3 \!\times\! 10^{-4}$ & 0.5 &  0.6 & 6.6 \\
         MnO \!\cite{PEPY1974433} & \hspace{0.05em} $2.5 \!\times\! 10^{-5}$ & 4.2 & 0.5 & 5.2\\
         Cr$_2$O$_3$ \!\cite{SAMUELSEN19691043} & \hspace{0.05em} $3.5 \!\times\! 10^{-5}$ & 0.3 & 0.9 & 4.9
    \end{tabular}
    \caption{Coefficients for the anti-ferromagnets considered here. $v_\theta$ is taken from the dispersion relation, $c_1$ is matched from neutron scattering data (Appendix~\ref{app:neutron}), and $\Lambda_{\rm UV}$ is estimated as the momentum for which the dispersion relation deviates from linear by $10\%$. The densities, $\rho_{\rm T}$, are taken from~\cite{doi:10.1063/1.4812323}.
    }
    \label{tab:coeffs}
\end{table}

\subsection{Dark matter--magnon interaction} \label{sec:DMmagnon}

We now study how a dark matter particle couples to the magnon modes introduced in the previous section. To do that, one starts from a specific model for the interaction of dark matter with the Standard Model. This is then computed in the non-relativistic limit, and matched with low-energy quantities, as we now show. For concreteness, we focus on two well motivated models, which serve as benchmarks to our general point. These were also studied in the context of ferromagnets~\cite{Trickle:2019ovy,Trickle:2020oki}. They are the magnetic dipole (m.d.) and the pseudo-mediated (p.m.) dark matter, which interact with the Standard Model electron respectively as~\cite[e.g.,][]{Sigurdson:2004zp,Masso:2009mu,Chang:2010en,Barger:2010gv,Fitzpatrick:2012ix,Gresham:2014vja,DelNobile:2014eta,Kavanagh:2018xeh,Chu:2018qrm,Banks:2010eh,Bagnasco:1993st,Chang:2009yt},
\begin{subequations} \label{eq:models}
    \begin{align}
        \mathcal{L}_{\chi}^{\rm m.d.} ={}& \frac{g_\chi}{\Lambda_\chi} V_{\mu\nu} \bar \chi \sigma^{\mu\nu} \chi + g_e V_\mu \bar e \gamma^\mu e \,, \\
        \mathcal{L}_{\chi}^{\rm p.m.} ={}& g_\chi \phi \, \bar \chi \chi + g_e \phi \, i \, \bar e  \gamma^5 e \label{eq:Lpm} \,,
    \end{align}
\end{subequations}
where $\phi$ and $V_\mu$ are ultra-light vector and pseudo-scalar mediators, $\chi$ and $e$ are the dark matter and electron fields, and $\Lambda_\chi$ is a UV scale pertaining to the dark sector. Moreover, $V_{\mu\nu} = \partial_\mu V_\nu - \partial_\nu V_\mu$ and $\sigma^{\mu\nu} = [\gamma^\mu, \gamma^\nu]$. The dipole models can naturally arise in certain technicolor theories, where the dark matter is a composite particle and can interact with the Standard Model through a vector mediator such as the dark photon \cite{Banks:2010eh,Bagnasco:1993st}. Similarly, pseudo-mediated models can dominate the spin response when the mediator is a pseudo-Goldstone boson \cite{Chang:2009yt}. 

To compute the dark matter--magnon interaction one can integrate out the mediator and perform the non-relativistic limit for both the dark matter and electron fields. This can be done either at the level of the matrix elements or integrating out anti-particles, similarly to the Heavy Quark Effective Theory procedure~\cite[e.g.,][]{Manohar:2000dt}. For the interested reader, we review this in Appendix~\ref{app:NRlimit}. After this, the dark matter--magnon interaction in the two instances is,
\begin{widetext}
    \begin{subequations} \label{eq:Lint}
        \begin{align}
            \begin{split}
                \mathcal{L}_{\rm int}^{\rm m.d.} ={}&  -\frac{4g_\chi g_e}{\Lambda_{\chi}m_e} \left( \chi^{\dagger}_{\rm nr} \frac{\sigma^i}{2} \chi_{\rm nr}\right) \left( \delta^{ij}-\frac{\nabla^{i}\nabla^j}{\nabla^2}\right) \left(  e^{\dagger}_{\rm nr} \frac{\sigma^j}{2} e_{\rm nr} \right) \xrightarrow[]{\; \text{IR} \;} - \frac{4g_\chi g_e}{\Lambda_{\chi} m_e} \left( \chi^{\dagger}_{\rm nr} \frac{\sigma^i}{2} \chi_{\rm nr}\right) \left( \delta^{ij}-\frac{\nabla^{i}\nabla^j}{\nabla^2}\right) {s}^j \,, 
            \end{split} \\
            \begin{split}
                \mathcal{L}_{\rm int}^{\rm p.m.} ={}& - \frac{g_\chi g_e}{m_e}  \chi^{\dagger}_{\rm nr} \chi_{\rm nr} \nabla^{-2} \bm{\nabla}\cdot\left( e^{\dagger}_{\rm nr} \frac{\bm \sigma}{2} e_{\rm nr} \right) \xrightarrow[]{\; \text{IR} \;} - \frac{g_\chi g_e}{m_e} \chi^{\dagger}_{\rm nr} \chi_{\rm nr} \nabla^{-2} \bm{\nabla}\cdot\bm{s} \,,
            \end{split}
        \end{align}
    \end{subequations}
\end{widetext}
where $\chi_{\rm nr}$ and $e_{\rm nr}$ are non-relativistic fields, and $\bm \sigma$ are Pauli matrices. We also used the fact that $e^\dagger_{\rm nr} \bm \sigma e_{\rm nr}/2$ is the electron spin density operator. When running towards low-energies, it will remain such, except that it must be expressed in terms of the correct low-energy degrees of freedom: the magnons rather than the single electrons.

One can now understand why anti-ferromagnets allow for multi-magnon emission while ferromagnets do not. As shown in Eqs.~\eqref{eq:Lint}, dark matter interacts with magnons via the spin density, whose components, $(s_x \pm i  s_y , s_z)$, have at most charge 1 under the unbroken $SO(2)$. In a ferromagnet this charge can be carried only by a single magnon mode. In an anti-ferromagnet, instead, there are two magnon modes carrying opposite charges. Hence any coupling to the spin density operator will allow multi-magnon emission.

Given the Lagrangians in Eqs.~\eqref{eq:Lint} and the spin density in Eq.~\eqref{eq:s}, one derives Feynman rules for the dark matter--magnon interaction, obtaining, 
\begin{align*}
    \begin{tikzpicture}
        [decoration={markings, mark = at position 0.68 with {\arrow{stealth}}}, baseline = 1.1em]
        \draw [postaction={decorate},thick] (0,0) -- (0.7,0); 
        \draw [postaction={decorate},thick] (0.7,0) -- (1.4,0); 
        \draw [postaction={decorate},dashed,thick] (0.7,0) -- (0.7,0.75);
        \node at (0.7,0) [circle,fill,inner sep=1.1pt]{};
        \node at (0.7,0.95) {\footnotesize $a, \lambda_1$};
        \node at (0.1,0.15) {\footnotesize $s$};
        \node at (1.3,0.2) {\footnotesize $s^\prime$};
    \end{tikzpicture} 
    ={}& - \frac{g_\chi g_e \sqrt{c_1}}{m_e} \omega \times 
    \begin{cases}
         \frac{4}{\Lambda_{\chi}} P_{ia}(\bm q) \sigma^i & \text{m.d.} \\
         {q}^a / {q}^2 & \text{p.m.}
    \end{cases} \,, \\[0.5em]
    \begin{tikzpicture}
        [decoration={markings, mark = at position 0.68 with {\arrow{stealth}}}, baseline = 0.8em]
        \draw [postaction={decorate},thick] (0,0) -- (0.7,0); 
        \draw [postaction={decorate},thick] (0.7,0) -- (1.4,0); 
        \draw [postaction={decorate},dashed,thick] (0.7,0) -- (0.3,0.75);
        \draw [postaction={decorate},dashed,thick] (0.7,0) -- (1.1,0.75);
        \node at (0.7,0) [circle,fill,inner sep=1.1pt]{};
        \node at (0.3,0.95) {\footnotesize $a,\lambda_1$};
        \node at (1.1,0.95) {\footnotesize $b,\lambda_2$};
        \node at (0.1,0.15) {\footnotesize $s$};
        \node at (1.3,0.2) {\footnotesize $s^\prime$};
    \end{tikzpicture} 
    ={}& \frac{g_\chi g_e}{m_e} (\omega_1 - \omega_2) \epsilon_{ab} \times 
    \begin{cases}
         \frac{4 }{\Lambda_{\chi}} P_{iz}(\bm q) \sigma^i   & \text{m.d.} \\
         {q}^z/ q^2  & \text{p.m.}
    \end{cases} \,.
\end{align*}
Solid lines represent a dark matter with polarization $s^{(\prime)}$, and dashed ones represent magnons with momenta $\bm q_{1,2}$, energies $\omega_{1,2}$, polarizations $\lambda_{1,2}$, and carrying an index $a,b=1,2$. The total momentum and energy carried by the magnons are $\bm q$ and $\omega$, with $P_{ij}(\bm q) \equiv \delta_{ij} - q_i q_j / q^2$. External dark matter lines come with standard non-relativistic bi-spinors, while external magnon lines come with a polarization vector, $\hat{\bm \varepsilon}_{\pm}=(1,\pm i)/\sqrt{2}$~\cite{Pavaskar:2021pfo}.

With this at hand, one can compute matrix elements for the emission rate of any number of gapless magnons with simple diagrammatic methods, exactly as one would do for relativistic particles.
In particular, the matrix element for the emission of \emph{any} number of low energy magnons is completely fixed by symmetry and by a single effective coefficient, $c_1$. In a more traditional formulation, the computation of multi-magnon scattering is substantially complicated by the failure of the Holstein--Primakoff approach, which mandates for a more involved treatment~\cite[e.g.,][]{dyson1956general}.

%%%%%%%%%%%%%%%%%%%%%%%%%%%%%%%%%%%%%%%%%%%%%%%%%%%%

\section{Event rates}
We now have everything we need to compute the expected event rates for the emission of one and two magnons by a dark matter particle.
For a target material with density $\rho_{\rm T}$, the total event rate per unit target mass can be evaluated by averaging the magnon emission rate over the dark matter velocity distribution, $f(\bm v_{\chi} + \bm v_{\rm e})$:
\begin{align}\label{eventrate}
    R = \frac{\rho_{\chi}}{\rho_{\rm T} m_{\chi}}\int d^3 v_{\chi} f(\bm v_{\chi} + \bm v_{\rm e})  \Gamma(v_{\chi}) \,.
\end{align}
The local dark matter density is taken to be $\rho_{\chi}=0.4$~GeV/cm$^3$~\cite{Piffl:2013mla}. The velocity distribution in the Milky Way is instead considered as a truncated Maxwellian given by the standard halo model, with dispersion $v_0 = 230$~km/s, escape velocity $v_{\rm esc} = 600$~km/s and boosted with respect to the galactic rest frame by the Earth velocity, $v_{\rm e} = 240$~km/s~\cite{Piffl:2013mla,Monari:2018ckf}. 
In the following, we present the projected reach for the case of single and two-magnon emission for the anti-ferromagnets NiO, MnO and Cr$_2$O$_3$.

\subsection{One magnon}

Using the Feynman rules presented in Section~\ref{sec:DMmagnon}, we compute the rates for the emission of a single gapless magnon. For the two benchmark models they read,
\begin{align}\label{singlemagnonrates}
    \frac{d\Gamma}{d\omega} = \frac{g_\chi^2 g_e^2 c_1}{\pi v_\chi m_e^2} \times \begin{cases} \frac{1+ \langle\cos^2\!\eta \rangle }{2 v_{\theta}^2 \Lambda_{\chi}^2}  \omega^2 \hspace{0.1 cm}  &\text{m.d.} \\[0.5em]
	\frac{\langle\sin^2\!\eta\rangle}{4}  \hspace{0.1 cm}  &\text{p.m.}
	\end{cases}\,,
\end{align}
where $\eta$ is the angle between the N\'eel vector, $\bm{\mathcal{N}}$, and the magnon momentum, $\bm{q}$, and $\langle \, \cdots \rangle$ represents an average over the direction of the latter. The decay rate is then a function of the relative angle between the magnetization and the direction of the incoming dark matter. Moreover, the magnon is emitted at fixed Cherenkov angle with respect to the incoming dark matter, $\cos\theta = q/(2m_{\chi}v_{\chi}) + v_{\theta}/v_{\chi}$. The final event rate, Eq.~\eqref{eventrate}, depends on the relative angle between $\bm v_{\rm e}$ and $\bm{\mathcal{N}}$. This leads to a daily modulation, which can possibly be used for background discrimination. To reduce the computational burden, we fix the two vectors to be parallel. More details can be found in Appendix~\ref{Rate}.

To obtain the decay rate, one integrates Eq.~\eqref{singlemagnonrates} over magnon energies between $\omega_{\min}$ and $\omega_{\max}$. 
The first one is set by the detector energy threshold. For the case of calometric readout, the best sensitivities that have been envisioned are of $\mathcal{O}(\text{meV})$~\cite{Hochberg_2016}. We thus set $\omega_{\rm min}=1$~meV. The maximum magnon energy is instead set by either the cutoff of the EFT or by kinematics. Specifically, $\cos\theta<1$ limits the possible momentum transfer, implying $\omega_{\text{max}}=\text{min}(v_{\theta}\Lambda_{\rm UV},2m_{\chi}v_{\theta}(v_{\chi}-v_{\theta}))$.  

\begin{figure}
    \centering
    \includegraphics[width=\columnwidth]{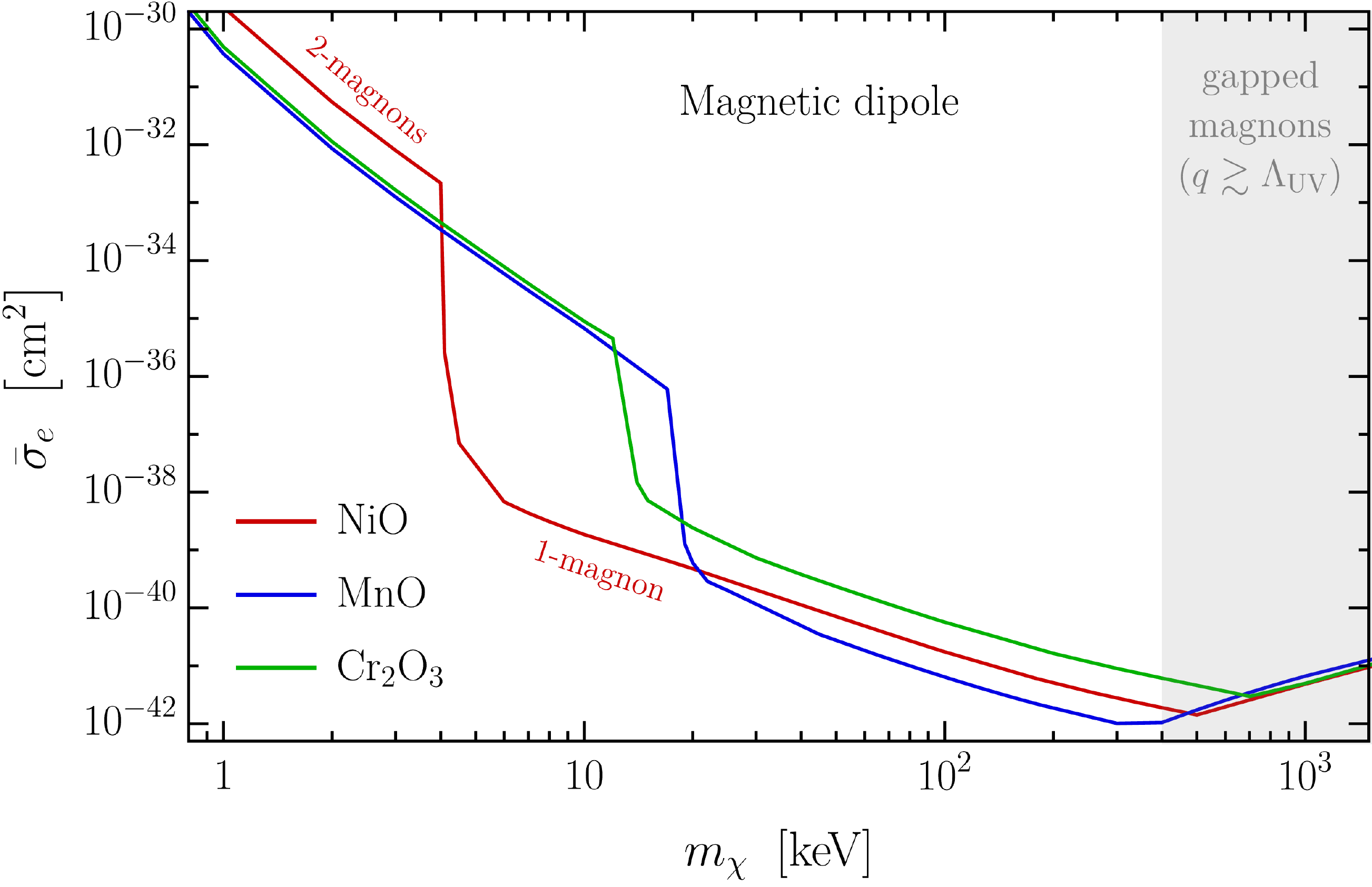}
    \includegraphics[width=\columnwidth]{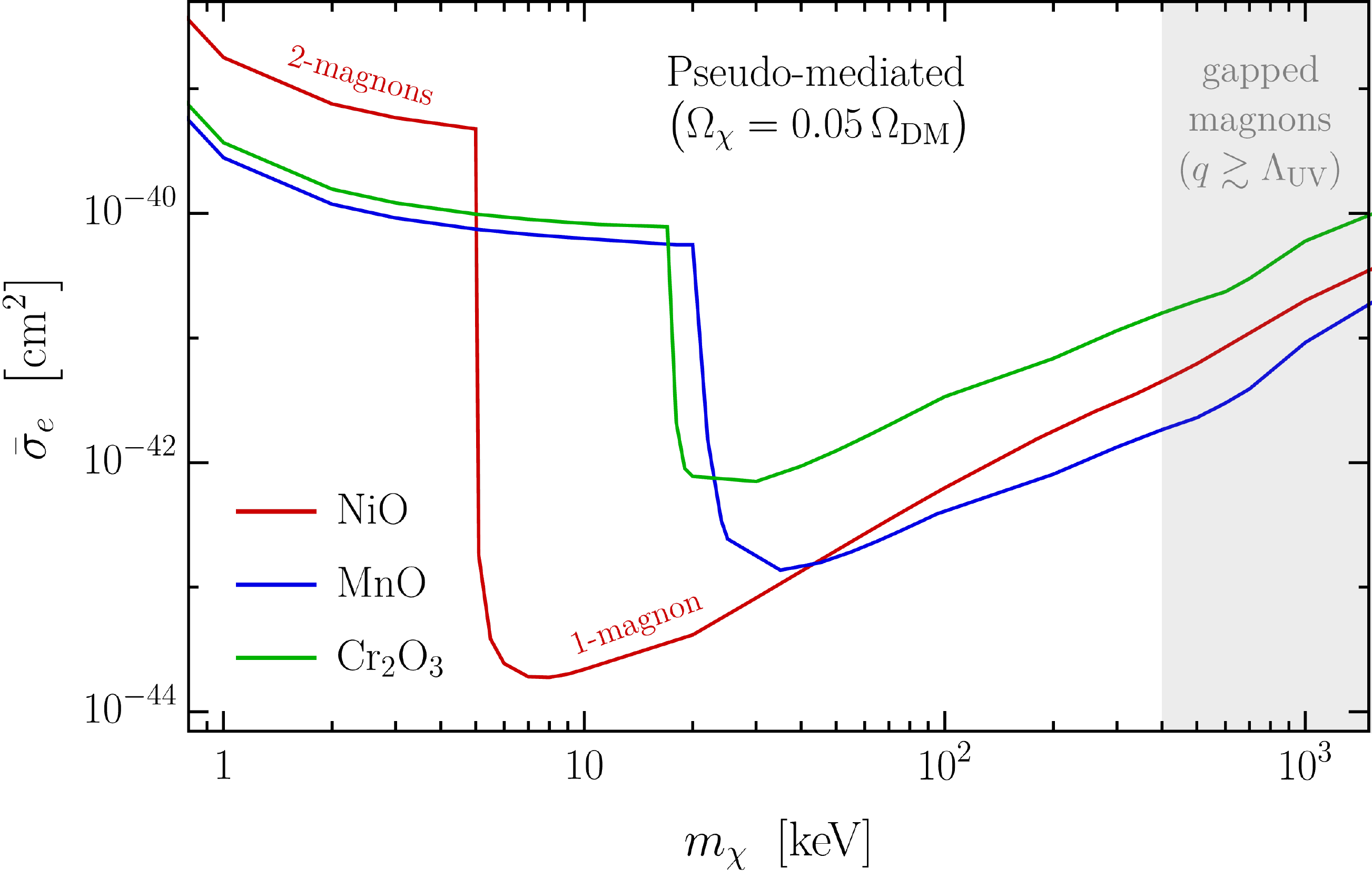}
    \caption{Projected reach at 95\% C.L. for a kilogram of material and a year of exposure assuming zero background, for the magnetic dipole ({\bf upper panel}) and pseudo-mediated ({\bf lower panel}) models. For the latter we assume $\Omega_\chi / \Omega_{\rm DM} = 0.05$. The lowest mass region is reached via the two-magnon channel. The gray region corresponds to masses for which gapped magnons are expected to play an important role. The magnetization is taken to be parallel to the Earth's velocity.}
    \label{fig:reach}
\end{figure}

Our projected reach for the three target materials are shown in Figure~\ref{fig:reach}, as compared to the following dark matter--electron cross sections, obtained from the interactions in Eqs.~\eqref{eq:models} in vacuum evaluated at the reference momentum $q_0=\alpha \, m_e$~\cite{Trickle:2019ovy}:
\begin{align}
    \bar \sigma_e =  \frac{g_{\chi}^2 g_e^2}{\pi} \times \begin{cases} \frac{1}{ \Lambda_{\chi}^2} \frac{6m_{\chi}^2+ m_e^2}{(m_{\chi}+m_e)^2}\hspace{0.1 cm}  &\text{m.d.} \\[0.5em]
	\frac{1}{4 \alpha^2 m_e^2} \frac{m_{\chi}^2}{(m_{\chi}+m_e)^2 }
  \hspace{0.1 cm}  &\text{p.m.}
	\end{cases} \,.
\end{align}
%
%\ang{Moreover, to avoid white dwarf cooling and self-interacting dark matter constraints, we impose for the $\chi$ in the pseudo-mediated model to be a $5\%$ sub-component of dark matter~\cite{Trickle:2019ovy}.}. 
Moreover, to avoid white dwarf cooling and self-interacting dark matter constraints, we impose $\chi$ to be a $5\%$ sub-component of dark matter for the pseudo-mediated model~\cite{Trickle:2019ovy}. Following convention, and for a simpler comparison with other proposals, we also assume zero background.

Importantly, NiO is sensitive to masses down to $m_\chi \simeq 5$~keV, even in the single magnon channel. This, as mentioned in the Introduction, is due to the good matching between the magnon and dark matter velocities. 
For $m_\chi \gtrsim 1 \text{ MeV}$, the rate starts receiving contributions from momenta above the cutoff, indicating that gapped magnons, not captured by the EFT, become relevant.

\subsection{Two magnons}

We consider now the emission of two magnons of energies and momenta $\omega_{1,2}$ and $\bm{q}_{1,2}$, and total energy and momentum $\omega$ and $\bm q$, by a dark matter of initial and final momentum $\bm k$ and $\bm k^\prime$. Using conservation of energy and momentum, the decay rate can be written as
\begin{align}
    \begin{split}
        % \Gamma ={}& \frac{1}{4 (2\pi)^3 v_{\theta}^3 }\int \frac{k^{\prime 2} dk^\prime \, d \text{cos}\theta \, dq_1}{q} |\mathcal{M}|^2 \\
        % ={}& \frac{1}{4 (2\pi)^3 v_{\theta}^3 v_{\chi} } \int d\omega \, dq \, dq_1 |\mathcal{M}|^2 \,,
          \Gamma ={}& \!\int \frac{k^{\prime 2} dk^\prime  d \text{cos}\psi  dq_1}{4 (2\pi)^3 v_{\theta}^3 \, q} |\mathcal{M}|^2 = \!\int \frac{d\omega  dq  dq_1}{4 (2\pi)^3 v_{\theta}^3 v_{\chi}} |\mathcal{M}|^2 \,,
    \end{split}
\end{align}
were $\psi$ denotes the angle between $\bm k$ and $\bm k^\prime$. In the second equality we also used $k^{\prime 2} dk^\prime \, d\cos\psi = (q/v_\chi) dq \, d\omega$.
Conservation of energy and momentum further implies $q_1 \leq (\omega + v_\theta q)/(2v_\theta)$. Thus, including also the EFT cutoff, the integral above is performed over $\omega_{\rm min} \leq \omega \leq \min\big(v_\theta\Lambda_{\rm UV}, \frac{1}{2}m_\chi v_\chi^2\big)$, $0 \leq q \leq \min\big(\Lambda_{\rm UV}, 2m_\chi v_\chi\big)$, and $0 \leq q_1 \leq \min\big(\Lambda_{\rm UV}, (\omega + v_\theta q)/(2v_\theta)\big)$. Again, we assume ideal calorimetric readout and set $\omega_{\rm min} = 1$~meV.

The projected reach for the two-magnon case is again shown in Figure~\ref{fig:reach}. This process allows to explore an even larger parameter space, by going down to masses as low as $m_\chi \sim 1$~keV. Due to improved kinematic matching, the mass reach is now almost independent of the target material. 
For the lightest dark matter, the momentum transfer is much smaller than the energy transfer, and the two magnons are emitted almost back-to-back, with possible interesting implications for background rejection~\cite{Caputo:2020sys}.
Similar to the one magnon case, the EFT predictions are unreliable for masses above 1 MeV.

%%%%%%%%%%%%%%%%%%%%%%%%%%%%%%%%%%%%%%%%%%%%%%%%%%%%

\section{Conclusion}

We have shown how well-assessed anti-ferromagnets can be used as optimal probes for sub-MeV dark matter with spin-dependent interactions. At low energies, these materials feature gapless magnons with two different polarizations, hence allowing for the emission of an arbitrary number of excitations. This, in turns, extends the potential reach down to $m_\chi \sim \mathcal{O}(\text{keV})$. As compared to ferromagnets~\cite{Trickle:2019ovy}, they have similar sensitivites on the dark matter couplings, but they probe masses more than an order of magnitude lighter. Interestingly, one of these anti-ferromagnets, nickel oxide, sustains magnon modes with a propagation speed accidentally close to the typical dark matter velocity, which allows it to absorb most of the dark matter energy already via the (dominant) one-magnon channel.
These results complements what already proposed for dark matter with spin-independent interactions~\cite[e.g.,][]{Schutz:2016tid,Knapen:2016cue,Acanfora:2019con,Knapen:2017ekk,Campbell-Deem:2019hdx}, allowing to cover the same mass region.

Moreover, the introduction of an EFT treatment to the problem opens the possibility for the evaluation of more involved observables as, for example, multi-magnon events with strong directionality and the potential for background discrimination~\cite[e.g.,][]{Caputo:2020sys}. Finally, a simple extension of our EFT would allow the description of the magnon--phonon coupling~\cite{Pavaskar:2021pfo}. 
This could be used to probe both spin-dependent and spin-independent interactions with a single target. Finally, it is interesting to study the prospect of our materials for axion absorption, and compare with the proposal in~\cite{Kakhidze:1990in,Marsh:2018dlj,Mitridate:2020kly,Chigusa:2020gfs,Schutte-Engel:2021bqm} (see also,~\cite{Barbieri:2016vwg,Crescini:2018qrz,QUAX:2020adt}). We leave these and other questions for future work.

%\ang{Broadly speaking, the dark matter theory space is so far largely unconstrained, leaving much room for possible properties of these particles. In light of this, it is important to try to undergo a wide, agnostic exploration. We showed that anti-ferromagnets (most notably, nickel-oxide) are likely optimal targets to be exploited in this program, specifically when looking for dark matter particles with masses in the keV to MeV range and spin-dependent interactions.}

%%%%%%%%%%%%%%%%%%%%%%%%%%%%%%%%%%%%%%%%%%%%%%%%%%%%

\begin{acknowledgments}
    We are grateful to Kim~Berghaus, Andrea~Caputo, Gianluca~Cavoto, Roberta~Citro, Andrea~Mitridate ,
    Riccardo~Penco and Ira~Z.~Rothstein for useful discussions and comments on the manuscript. For most of the development of this work A.E. has been a Roger Dashen Member at the Institute for Advanced Study, whose work was also supported by the U.S. Department of Energy, Office of Science, Office of High Energy Physics under Award No. DE-SC0009988. 
\end{acknowledgments}

%%%%%%%%%%%%%%%%%%%%%%%%%%%%%%%%%%%%%%%%%%%%%%%%%%%%

\appendix 

\section{Matching with neutron scattering} \label{app:neutron}
To determine the coefficients $c_1$ reported in Table~\ref{tab:coeffs}, one has to perform a matching calculation between the EFT and a short distance theory. To do this, we compute the one-magnon neutron scattering cross section both within the EFT and within the microscopic theory, which we take to be the Heisenberg model (see for, example,~\cite{altland_simons_2010} for a pedagogical discussion). This will allow us to extract $c_1$ in terms of the exchange constants in the Heisenberg model and of the magnon velocities, both of which have been measured in neutron scattering experiments. 

We begin with the calculation in the EFT. In a neutron scattering experiment, the interaction between the neutron and the spins of the magnetic material is mediated by an off-shell photon.
This is because both the neutron, $n$, and the spins, $\bm{\mathcal{S}}_i$, couple to the magnetic field via the Zeeman coupling~\cite[e.g.,][]{Burgess:1998ku}, corresponding to an interaction Lagrangian given by
\begin{align}
    \begin{split}
        \mathcal{L}_{ns} ={}& - \gamma \mu_{n
        } n^{\dagger} \bm{\sigma} \, n \cdot \bm{B} - g \mu_{\rm B} \sum_i \bm{B} \cdot \bm{\mathcal{S}}_i   \\  
        \xrightarrow[]{\; \text{IR} \;}{}& - \gamma \mu_{n} n^{\dagger} \bm{\sigma} \, n \cdot \bm{B} - g \mu_{\rm B} \bm{B} \cdot \bm{s} \,,
    \end{split}
\end{align}
where $\mu_{\rm B} = e / (2 m_e)$ is the Bohr magneton, $\mu_n = (m_e/m_n) \mu_{\rm B}$ is the neutron magnetic moment, and the two g-factors are given by $\gamma \simeq 1.9$ and $g \simeq 2$.
In going to the second line, we have taken the IR limit where the effective coupling of the magnetic field to the spins is via the spin density. Integrating out the gauge field results in an effective interaction between the neutron and the magnons appearing in the spin density operator. This interaction is given by 
\begin{align}
    \mathcal{L}_{ns} = g \mu_B \gamma \mu_n  \, \big(n^\dagger \bm \sigma \, n\big)  \cdot \nabla^{-2} \big( {\bm \nabla} \times ({\bm \nabla} \times \bm{s}) \big) \,.
\end{align}
We are interested in the one-magnon cross section due to neutron scattering. The Feynman rule for this interaction can be computed straightforwardly using the expression for the spin density in Eq.~\eqref{eq:s}:
\begin{align*}
    \begin{tikzpicture}
        [decoration={markings, mark = at position 0.68 with {\arrow{stealth}}}, baseline = 1.1em]
        \draw [postaction={decorate},thick] (0,0) -- (0.7,0); 
        \draw [postaction={decorate},thick] (0.7,0) -- (1.4,0); 
        \draw [postaction={decorate},dashed,thick] (0.7,0) -- (0.7,0.75);
        \node at (0.7,0) [circle,fill,inner sep=1.1pt]{};
        \node at (0.7,0.95) {\footnotesize $a, \lambda_1$};
        \node at (0.1,0.15) {\footnotesize $s$};
        \node at (1.3,0.2) {\footnotesize $s^\prime$};
    \end{tikzpicture} 
    ={}& - \gamma \mu_n g \mu_{\rm B} \sqrt{c_1} \omega  P_{ia}(\bm q) \sigma^i \,,
\end{align*}
where the solid lines represent the neutron with polarization $s(s')$ and the dashed ones represent the magnon carrying an index $a$, with momenta $\bm q$ and polarization $\lambda_1$. Taking the incoming and outgoing neutron energy-momenta to be $(E,\bm k)$ and $(E',\bm k')$ respectively, one can calculate the differential cross section to be
\begin{align} \label{eftCS}
    \frac{d^2\sigma}{d\Omega \, d E'} ={}& \frac{V }{(2\pi)^2} \frac{m_n k'}{v_n} (\gamma \mu_n g \mu_{\rm B})^2 c_1  \frac{1+\hat{q}_z^2}{4} \omega(q) \notag \\
    & \times \delta\big(E'-E-\omega(q)\big)  \\ ={}& V (\gamma r_0)^2 \frac{k'}{k}  c_1  \frac{1+\hat{q}_z^2}{4} \omega(q) \delta\big(E'-E-\omega(q)\big)  \notag
\end{align}
where we have rewritten the couplings in terms of the classical electron radius, $r_0 \equiv \frac{1}{4\pi} \frac{e^2}{m_e}$. Also, $V$ is the volume of the sample and $v_n$ represents the velocity of the incoming neutron. Since the energy of the outgoing neutron is the one that is measured, the differential rate has been calculated with respect to $E'$.   

Let us now compute this same cross section in the the Heisenberg model which describes the interaction between spins on a lattice, and which is typically employed to report experimental data. We will mostly follow the derivation and notation reported in~\cite{lovesey1984theory}. The Heisenberg model can be thought of as an effective theory of the Hubbard model which describes the interplay of electrons on a lattice \cite{altland_simons_2010}. 
For an anti-ferromagnet, one can think of the dynamics being described by two sublattices, where the spins belonging to the same sublattice point in the same direction, and the spins belonging to different sublattices point in opposite directions. The Heisenberg Hamiltonian for an anti-ferromagnet can then be written as,
\begin{align} \label{eq:HHeisenberg}
    \begin{split}
        H ={}& \sum_{\bm x, \bm R } J(R) \: \bm{\mathcal{S}}_{\bm x} \cdot \bm{\mathcal{S}}_{\bm x+\bm R} \\
        & + \sum_{\bm x,\bm r} J'(r) \: \bm{\mathcal{S}}_{\bm x} \cdot \bm{\mathcal{S}}_{\bm x+\bm r} \,.
    \end{split}
\end{align}
Here $\bm x$ is any position on the lattice, $\bm R$ is a vector connecting two sites on opposite sublattices, and $\bm r$ is a vector connecting two sites on the same sublattice.
Hence, $J(J')$ describes the exchange interaction between the spins on opposite (same) sublattices. From henceforth, we will work with a cubic lattice, which is sufficient for our discussion and often represents a good approximation of a realistic material. We can define the Fourier space analogs of these couplings as
\begin{subequations} \label{eq:FourierJ}
    \begin{align}
        \mathcal{J}(q) \equiv{}& \sum_{\bm R} J(R) e^{-i\bm q \cdot \bm R } \simeq \mathcal{J}_{(0)} - \frac{q^2}{6} \mathcal{J}_{(2)} \,, \\
        \mathcal{J}'(q) \equiv{}& \sum_{\bm r} J'(r) e^{-i\bm q \cdot \bm r } \simeq \mathcal{J}'_{(0)} - \frac{q^2}{6} \mathcal{J}'_{(2)} \,,
    \end{align}
\end{subequations}
where we have expanded in $q$ since we will be interested only in the long wavelength excitations. We also define $ \mathcal{J}_{(n)} \equiv \sum_{\bm R} J(R) R^n$ and similarly for $\mathcal{J}'$. Notice that the odd powers vanish due to the symmetry of the cubic lattice, which implies that the sum of every subset of sublattice vectors with fixed magnitude vanishes, $\sum_{R={\rm fix}}\bm R = 0$. 

Neglecting lattice vibrations, the differential cross section for inelastic neutron scattering in the Heisenberg model at zero temperature is given by~\cite{lovesey1984theory}
\begin{align}\label{heisenbergCS}
    \frac{d^2\sigma}{d\Omega \: dE'} =  (\gamma r_0)^2 \frac{k'}{k} |F(\bm q)|^2 (\delta_{ij}-\hat{q}_i\hat{q}_j) S^{ij}(\omega,\bm q) \,,
\end{align}
where $\bm q= \bm k'-\bm k$ and $\omega = E^\prime - E$ are respectively the momentum and energy transfer. Moreover, $F(\bm q)$ is the neutron form factor, normalized such that $F(0)=1$. The van-Hove scattering function,  $S^{ij}(\omega,\bm{q})$, is the Fourier transform of the spin--spin correlator:
\begin{align}\label{correlator}
    S^{ij}(\omega,\bm{q}) = \int \frac{dt}{2\pi} \sum_{\bm r}  e^{i\bm q\cdot \bm r-i \omega t} \langle  S^i(\bm r,t)  S^j(0,0) \rangle \,.
\end{align}
For case where the N\'eel vector of the anti-ferromagnet points in the $z$-direction, the only relevant spin correlators are the diagonal ones in the transverse directions. For a two-sublattice anti-ferromagnet, one can obtain the scattering function \cite{lovesey1984theory} to be 
\begin{align}
    S^{ab}(\omega,\bm{q}) = \frac{\delta^{ab}}{4} \delta\big(\omega-\omega(q)\big) \frac{ \mathcal{S}^2 N}{3v_{\theta}^2} \big(\mathcal{J}_{(2)} - \mathcal{J}'_{(2)}\big) \omega \,,
\end{align}
where $N$ is the number of unit cells in the sample and $\mathcal{S}$ is the magnitude of the spins. Now we are in position to compare the cross section computed within the EFT, Eq.~\eqref{eftCS}, and within the Heisenberg model, Eq.~\eqref{heisenbergCS}, using the above result. Since we are interested in the long-wavelength limit, we can set the neutron form factor to be unity. Matching the two computations, $c_1$ reads
\begin{align}
    c_1 = \frac{\mathcal{S}^2}{3 V_0 v_{\theta}^2} \big(\mathcal{J}_{(2)} - \mathcal{J}'_{(2)}\big) \,,
\end{align}
where $V_0=V/N$ is the volume of a unit cell. For a cubic lattice $V_0=a_0^3$, where $a_0$ is the lattice spacing of the crystal. 

\begin{figure}
    \centering
    \begin{tikzpicture}
        % solid edges
        \draw[] (-3,0) -- (3,0);
        \draw[] (0,0) -- (0,3);
        \draw[] (-3,3) -- (3,3);
        \draw[] (3,0) -- (3,3);
        \draw[] (-3,0) -- (-3,3);
        \draw[] (3,0) -- (4,0.7);
        \draw[] (3,3) -- (4,3.7);
        \draw[] (0,3) -- (1,3.7);
        \draw[] (-3,3) -- (-2,3.7);
        \draw[] (-2,3.7) -- (4,3.7);
        \draw[] (4,0.7) -- (4,3.7);
        
        % dashed edges
        \draw[dashed] (-3,0) -- (-2,0.7);
        \draw[dashed] (-2,0.7) -- (4,0.7);
        \draw[dashed] (-2,0.7) -- (-2,3.7);
        \draw[dashed] (0,0) -- (1,0.7);
        \draw[dashed] (1,0.7) -- (1,3.7);
        
        % dotted diagonals
        \draw[dotted,gray] (0,0) -- (-3,3);
        \draw[dotted,gray] (-3,0) -- (0,3);
        \draw[dotted,gray] (-3,0) -- (-2,3.7);
        \draw[dotted,gray] (-2,0.7) -- (-3,3);
        \draw[dotted,gray] (-2,0.7) -- (1,3.7);
        \draw[dotted,gray] (1,0.7) -- (-2,3.7);
        \draw[dotted,gray] (0,0) -- (-2,0.7);
        \draw[dotted,gray] (-3,0) -- (1,0.7);
        \draw[dotted,gray] (-3,3) -- (1,3.7);
        \draw[dotted,gray] (0,3) -- (-2,3.7);
        \draw[dotted,gray] (0,0) -- (1,3.7);
        \draw[dotted,gray] (1,0.7) -- (0,3);
        \draw[dotted,gray] (0,0) -- (3,3);
        \draw[dotted,gray] (3,0) -- (0,3);
        \draw[dotted,gray] (1,0.7) -- (4,3.7);
        \draw[dotted,gray] (4,0.7) -- (1,3.7);
        \draw[dotted,gray] (0,0) -- (4,0.7);
        \draw[dotted,gray] (3,0) -- (1,0.7);
        \draw[dotted,gray] (0,3) -- (4,3.7);
        \draw[dotted,gray] (3,3) -- (1,3.7);
        \draw[dotted,gray] (3,0) -- (4,3.7);
        \draw[dotted,gray] (4,0.7) -- (3,3);
        
        % filled sites
        \node at (0,0) [circle,fill,blue,inner sep=2pt]{};
        \node at (-2,0.7) [circle,fill,blue,inner sep=2pt]{};
        \node at (4,0.7) [circle,fill,blue,inner sep=2pt]{};
        \node at (3,3) [circle,fill,blue,inner sep=2pt]{};
        \node at (1,3.7) [circle,fill,blue,inner sep=2pt]{};
        \node at (-3,3) [circle,fill,blue,inner sep=2pt]{};
        \node at (-1.5,1.5) [circle,fill,blue,inner sep=2pt]{};
        \node at (-1,3.35) [circle,fill,blue,inner sep=2pt]{};
        \node at (0.5,1.85) [circle,fill,blue,inner sep=2pt]{};
        \node at (2,0.35) [circle,fill,blue,inner sep=2pt]{};
        \node at (2.5,2.2) [circle,fill,blue,inner sep=2pt]{};
        
        % empty sites
        \node at (0,3) [circle,draw,blue,inner sep=2pt,thick]{};
        \node at (1,0.7) [circle,draw,blue,inner sep=2pt,thick]{};
        \node at (1.5,1.5) [circle,draw,blue,inner sep=2pt,thick]{};
        \node at (3,0) [circle,draw,blue,inner sep=2pt,thick]{};
        \node at (2,3.35) [circle,draw,blue,inner sep=2pt,thick]{};
        \node at (4,3.7) [circle,draw,blue,inner sep=2pt,thick]{};
        \node at (3.5,1.85) [circle,draw,blue,inner sep=2pt,thick]{};
        \node at (-1,0.35) [circle,draw,blue,inner sep=2pt,thick]{};
        \node at (-3,0) [circle,draw,blue,inner sep=2pt,thick]{};
        \node at (-2.5,1.85) [circle,draw,blue,inner sep=2pt,thick]{};
        \node at (-2,3.7) [circle,draw,blue,inner sep=2pt,thick]{};
        \node at (-0.5,2.2) [circle,draw,blue,inner sep=2pt,thick]{};
        
        % sublattice vectors
        \draw[line width=1.2pt,YellowOrange,->] (0,0) -- (-3,0);
        \draw[line width=1.2pt,YellowOrange,->] (0,0) -- (-0.5,2.2);
        \draw[line width=1.2pt,YellowOrange,->] (0,0) -- (1.5,1.5);
        \draw[line width=1.2pt,Maroon,->] (0,0) -- (-1.5,1.5);
        \draw[line width=1.2pt,Maroon,->] (0,0) -- (1,3.7);
        \draw[line width=1.2pt,Maroon,->] (0,0) -- (2.5,2.2);
        
        % vector labels
        \node at (-1.5,-0.25) {\footnotesize \color{YellowOrange} $\bm R_2$};
        \node at (-0.55,1.3) {\footnotesize \color{YellowOrange} $\bm R_3$};
        \node at (1.,1.35) {\footnotesize \color{YellowOrange} $\bm R_1$};
        \node at (-1.15,0.9) {\footnotesize \color{Maroon} $\bm r_1$};
        \node at (2.3,1.7) {\footnotesize \color{Maroon} $\bm r_2$};
        \node at (0.5,2.6) {\footnotesize \color{Maroon} $\bm r_3$};
        
        % vertical lengths
        \draw[gray,<->] (-3.5,0) -- (-3.5,3);
        \node at (-3.7,1.5) {\footnotesize \color{Gray}  $a_0$};
        
    \end{tikzpicture}
    \caption{Crystal structure of the NiO anti-ferromagnets. Filled circles represent spins pointing in one direction, while empty circles represent spins pointing in the opposite direction.}
    \label{fig:NiOstructure}
\end{figure}
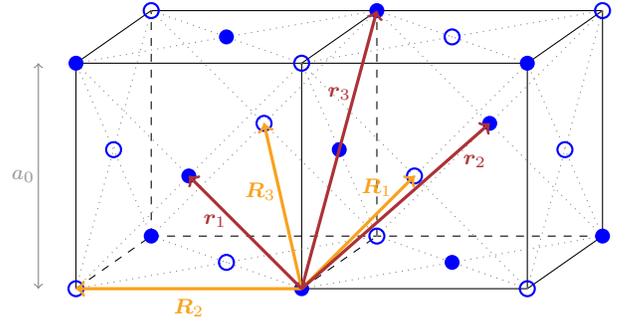

We now only need to specify $\mathcal{J}_{(2)}$ and $\mathcal{J}_{(2)}^\prime$ for a given material. We will show how to do that explicitly for NiO. The other anti-ferromagnets considered in this work follow similar procedures, albeit more tedious. 
The crystal structure of NiO is shown in Figure~\ref{fig:NiOstructure}. The corresponding couplings and crystal parameters have been measured in~\cite{hutchings_inelastic_1971,hutchings1972measurement}. From the definitions in Eqs.~\eqref{eq:FourierJ}, the Fourier space moments are found to be
\begin{align*}
    \begin{split}
        \mathcal{J}_{(2)} ={}& 6 R_1^2 J(R_1) + 6 R_2^2 J(R_2) + 12 R_3^2 J(R_3) + \dots \,, \\
        \mathcal{J}_{(2)}^\prime ={}& 6 r_1^2 J^\prime(r_1) + 12 r_2^2 J^\prime(r_2) + 12 r_3^2 J^\prime(r_3) + \dots \,,
    \end{split}
\end{align*}
where the numerical factors in front of each term correspond to the number of vectors with the same magnitude, as reported in~\cite{hutchings1972measurement}.\footnote{We point out that, in our notation of Eq.~\eqref{eq:HHeisenberg} each coupling $J^{(\prime)}$ is counted twice in the sum, while in the notation of~\cite{hutchings_inelastic_1971,hutchings1972measurement} only once. For a proper matching, therefore, one needs to recall that our couplings are half of those measured in~\cite{hutchings_inelastic_1971,hutchings1972measurement}.} From Figure~\ref{fig:NiOstructure} we also deduce the length of the different vectors, i.e.\footnote{The NiO crystal presents a small anisotropy~\cite{hutchings1972measurement}, which makes it deviate from a perfect cubic crystal. This also gives a tiny gap to lowest lying magnon modes. All these effects are negligible for our purposes.}
\begin{align}
    \begin{split}
        & R_1 = \frac{a_0}{\sqrt{2}} \,, \quad R_2 = a_0 \,, \quad R_3 = \sqrt{\frac{3}{2}} a_0 \,, \\
        & r_1 = \frac{a_0}{\sqrt{2}} \,, \quad r_2 = \sqrt{\frac{3}{2}}a_0 \,, \quad r_3 = \sqrt{2} a_0 \,.
    \end{split}
\end{align}
For the case of NiO, the next-to-nearest neighbor coupling $J(R_2) \simeq 9.5$~meV is much larger than the other exchange constants~\cite{hutchings_inelastic_1971,hutchings1972measurement}, and hence dominates the expressions above. We then have,
\begin{align}
     c_1 \simeq \frac{4\mathcal{S}^2}{a_0 v_{\theta}^2} J(R_2) \,.
\end{align}
For NiO, the measured parameters are~\cite{hutchings_inelastic_1971,hutchings1972measurement} $\mathcal{S}=1$, $a_0 \simeq 4.17$~\AA, and $v_{\theta} \simeq 1.3 \times 10^{-4}$. From which we obtain $c_1 \simeq 0.5$~MeV/\AA.

%%%%%%%%%%%%%%%%%%%%%%%%%%%%%%%%%%%%%%%%%%%%%%%%%%%

\section{Non-relativistic limit} \label{app:NRlimit}

To perform the non-relativistic limit at the level of the Lagrangian, one can follow a procedure analogous to what done in Heavy Quark Effective Theory~\cite[e.g.,][]{Manohar:2000dt}, and essentially amounting to integrating out the anti-particles and expanding in small velocities.

Let us consider the electron field as an example; the same procedure applies to any spin-1/2 field. Firstly, one starts from the relativistic 4-component Dirac field and performs the following splitting,
\begin{align} \label{eq:esplit}
    e = {\rm e}^{-i m_e t}\big( e_{\rm L} + e_{\rm H} \big) \,, \;\;\, \text{with} \;\,
    \begin{cases}
        e_{\rm L} = {\rm e}^{i m_e t} \frac{\mathbbm{1} + \gamma^0}{2} e \\
        e_{\rm H} = {\rm e}^{i m_e t} \frac{\mathbbm{1} - \gamma^0}{2} e
    \end{cases} \!\!\!\!\!,
\end{align}
where the $\gamma$-matrices are written in the Dirac basis. The projectors in the equation above are such that, in the non-relativistic limit, $e_{\rm L}$ only contains the two upper components (associated to the electron), while $e_{\rm H}$ only the two lower ones (associated to the positron). The overall phase ensures that $e_{\rm L}$ has only support on small energies, $\omega \sim k^2/(2m_e) \ll m_e$, while $e_{\rm H}$ contains large energies, $\omega \sim 2m_e$. Indeed, by plugging Eq.~\eqref{eq:esplit} in the Dirac Lagrangian one gets,
\begin{align}
    \begin{split}
        \mathcal{L}_e \simeq{}& - \bar{e}_{\rm H}\left( i \partial_t + 2m_e\right) e_{\rm H} + \bar e_{\rm L} \,i \partial_t e_{\rm L} \\
        & + \bar e_{\rm H} \,i \bm{\gamma}\cdot\bm{\nabla} e_{\rm L} + \bar e_{\rm L} \,i \bm{\gamma}\cdot\bm{\nabla}e_{\rm H} \,,
    \end{split}
\end{align}
where we already expanded in the non-relativistic limit.\footnote{At lowest order, this corresponds to taking the standard Heavy Quark Effective Theory Lagrangian~\cite{Manohar:2000dt} and set the 4-velocity to $v_\mu = (1,\bm 0)$.} From the Lagrangian above we see that, as anticipated, the particle field $e_{\rm L}$ is massless, while the anti-particle field $e_{\rm H}$ is massive, with mass $2m_e$. As such, it can be integrated out at low-energies. At tree level, the equations of motion give
\begin{align}
    e_{\rm H} \simeq \frac{i}{2m_e} \bm \gamma \cdot \bm \nabla e_{\rm L} \,.
\end{align}

We can now use this to take the non-relativistic limit of the interaction between dark matter and the Standard Model electron. We will work in the simplest case of pseudo-mediated dark matter. The magnetic dipole dark matter follows the same procedure, albeit slightly more tedious. The initial relativistic Lagrangian is the one in Eq.~\eqref{eq:Lpm}. Using the fact that, due to the projectors in Eq.~\eqref{eq:esplit}, $\bar e_{\rm H} e_{\rm L} = \bar e_{\rm H} \gamma^5 e_{\rm H} = \bar e_{\rm L} \gamma^5 e_{\rm L} = 0$, it is simple to show that,
\begin{align}
    \bar \chi \chi \simeq \bar \chi_{\rm L} \chi_{\rm L} \,, \quad \bar e \,i \gamma^5 e \simeq \frac{1}{2m_e} \bm \nabla \cdot \left( \bar e_{\rm L} \bm \gamma \gamma^5 e_{\rm L} \right) \,.
\end{align}
Moreover, one also has
\begingroup
    \renewcommand*{\arraystretch}{0.9}
    \begin{align}
        \bm \gamma \gamma^5 \frac{\mathbbm{1}+\gamma^0}{2} = \left(\begin{matrix} \bm \sigma & 0 \\ 0 & 0 \end{matrix}\right) \equiv \bm{\Sigma} \,,
    \end{align}
\endgroup
which is precisely the spin operator.
It then follow that, including the ultra-light mediator, the non-relativistic Lagrangian reads,
\begin{align}
    \mathcal{L}^{\rm p.m.}_\chi \simeq{}& \frac{1}{2}\big(\partial\phi\big)^2 + g_\chi \phi \, \bar \chi_{\rm L} \chi_{\rm L} + \frac{g_e}{m_e} \phi\, \bm\nabla \cdot\left( \bar e_{\rm L} \frac{\bm \Sigma}{2} e_{\rm L} \right) \notag \\
    \begin{split}
        \to{}& \frac{g_e g_\chi}{m_e} \bar \chi_{\rm L} \chi_{\rm L} \,\square^{-1} \bm\nabla \cdot\left( \bar e_{\rm L} \frac{\bm \Sigma}{2} e_{\rm L} \right) \\
        \simeq{}& -\frac{g_e g_\chi}{m_e} \bar \chi_{\rm L} \chi_{\rm L} \,\nabla^{-2} \bm\nabla \cdot\left( \bar e_{\rm L} \frac{\bm \Sigma}{2} e_{\rm L} \right)
    \end{split} \\
    ={}& -\frac{g_e g_\chi}{m_e} \chi_{\rm rn}^\dagger \chi_{\rm nr} \nabla^{-2} \bm\nabla\cdot\left( e_{\rm nr}^\dagger \frac{\bm \sigma}{2} e_{\rm nr} \right) \,, \notag
\end{align}
where we first integrated out the mediator and then performed the non-relativistic limit again. In the last line we switched to the two-component fields, related to the light field by $e_{\rm L} = {(e_{\rm nr} \; 0)}^{\rm t}$. 

%%%%%%%%%%%%%%%%%%%%%%%%%%%%%%%%%%%%%%%%%%%%%%%%%%%

\section{Computation of event rates} \label{Rate}
To compute the event rates, one calculates the convolution of the dark matter velocity distribution function and the decay rates as in Eq.~\eqref{eventrate}. The velocity distribution function boosted to the Earth's rest frame is given by 
\begin{align}
    f(\bm v_{\chi} + \bm v_{\rm e}) = \frac{1}{N_0} {\rm e}^{-\frac{(\bm v_{\chi} + \bm v_{\rm e})^2}{v_0^2}}\Theta(v_{\rm esc}-|\bm v_{\chi}+\bm v_{\rm e}|) \,, 
\end{align}
where $N_0 = \pi v_0^2 [\sqrt{\pi} v_0 \text{erf}(v_{\rm esc}/v_0) - 2 v_{\rm esc}\text{exp}(-v_{\rm esc}^2 /v_0^2)]$. Shifting the velocity to $\bm v_{\chi} \rightarrow \bm v_{\chi}-\bm v_{\rm e}$, one finds the total event rate,
\begin{align} \label{eq:R}
  R=  \frac{\rho_{\chi}}{\rho_{\rm T} m_{\chi} N_0}\int_{v_\chi \leq v_{\rm esc}} \!\!\!\! d^3 v_{\chi} {\rm e}^{-{v^2_{\chi}}/{v_0^2}} \Gamma(|\bm v_{\chi}-\bm v_{\rm e}|,\theta^\prime) \,,
\end{align}
where $\theta^\prime$ is the relative angle between the direction of the incoming dark matter, $\hat{\bm v}_\chi$, and the magnetization $\bm{\mathcal{N}}$. This is related to the angle between the momentum transfer and the magnetization, $\eta$, appearing in Eq.~\eqref{singlemagnonrates} by~\cite[e.g.,][]{Caputo:2020sys},
\begin{align}
\cos \eta=\cos\theta \cos\theta' + \sin\theta \sin\theta' \sin(\phi-\phi') \,,
\end{align}
with $(\theta',\phi')$ denoting the angle between the incoming dark matter and the magnetization vector and $(\theta,\phi)$ denoting the angle between the momentum transfer and the incoming dark matter. From momentum conservation, one obtains $\cos \theta = q/(2 m_{\chi} v_{\chi}) + \omega/(v_{\chi}q)$. When considered for single magnon emission (i.e., $\omega = v_\theta q$) this returns the usual Cherenkov condition.  

After integrating over the direction of $\bm v_\chi$ in Eq.~\eqref{eq:R}, the final event rate depends only on the relative angle between the Earth's velocity and the magnetization. Our results of Figure~\ref{fig:reach} are computed in the simplified case where $\hat{\bm v}_{\rm e} \cdot \hat{\bm{\mathcal{N}}}=1$. The integral in Eq.~\eqref{eq:R} is evaluated numerically.

\bibliographystyle{apsrev4-2}
\bibliography{biblio.bib}

\begin{thebibliography}{109}%
\makeatletter
\providecommand \@ifxundefined [1]{%
 \@ifx{#1\undefined}
}%
\providecommand \@ifnum [1]{%
 \ifnum #1\expandafter \@firstoftwo
 \else \expandafter \@secondoftwo
 \fi
}%
\providecommand \@ifx [1]{%
 \ifx #1\expandafter \@firstoftwo
 \else \expandafter \@secondoftwo
 \fi
}%
\providecommand \natexlab [1]{#1}%
\providecommand \enquote  [1]{``#1''}%
\providecommand \bibnamefont  [1]{#1}%
\providecommand \bibfnamefont [1]{#1}%
\providecommand \citenamefont [1]{#1}%
\providecommand \href@noop [0]{\@secondoftwo}%
\providecommand \href [0]{\begingroup \@sanitize@url \@href}%
\providecommand \@href[1]{\@@startlink{#1}\@@href}%
\providecommand \@@href[1]{\endgroup#1\@@endlink}%
\providecommand \@sanitize@url [0]{\catcode `\\12\catcode `\$12\catcode
  `\&12\catcode `\#12\catcode `\^12\catcode `\_12\catcode `\%12\relax}%
\providecommand \@@startlink[1]{}%
\providecommand \@@endlink[0]{}%
\providecommand \url  [0]{\begingroup\@sanitize@url \@url }%
\providecommand \@url [1]{\endgroup\@href {#1}{\urlprefix }}%
\providecommand \urlprefix  [0]{URL }%
\providecommand \Eprint [0]{\href }%
\providecommand \doibase [0]{https://doi.org/}%
\providecommand \selectlanguage [0]{\@gobble}%
\providecommand \bibinfo  [0]{\@secondoftwo}%
\providecommand \bibfield  [0]{\@secondoftwo}%
\providecommand \translation [1]{[#1]}%
\providecommand \BibitemOpen [0]{}%
\providecommand \bibitemStop [0]{}%
\providecommand \bibitemNoStop [0]{.\EOS\space}%
\providecommand \EOS [0]{\spacefactor3000\relax}%
\providecommand \BibitemShut  [1]{\csname bibitem#1\endcsname}%
\let\auto@bib@innerbib\@empty
%</preamble>
\bibitem [{\citenamefont {Agnese}\ \emph {et~al.}(2016)\citenamefont {Agnese}
  \emph {et~al.}}]{SuperCDMS:2015eex}%
  \BibitemOpen
  \bibfield  {author} {\bibinfo {author} {\bibfnamefont {R.}~\bibnamefont
  {Agnese}} \emph {et~al.} (\bibinfo {collaboration} {SuperCDMS}),\ }\href
  {https://doi.org/10.1103/PhysRevLett.116.071301} {\bibfield  {journal}
  {\bibinfo  {journal} {Phys. Rev. Lett.}\ }\textbf {\bibinfo {volume} {116}},\
  \bibinfo {pages} {071301} (\bibinfo {year} {2016})},\ \Eprint
  {https://arxiv.org/abs/1509.02448} {arXiv:1509.02448 [astro-ph.CO]}
  \BibitemShut {NoStop}%
\bibitem [{\citenamefont {Akerib}\ \emph {et~al.}(2017)\citenamefont {Akerib}
  \emph {et~al.}}]{LUX:2016ggv}%
  \BibitemOpen
  \bibfield  {author} {\bibinfo {author} {\bibfnamefont {D.~S.}\ \bibnamefont
  {Akerib}} \emph {et~al.} (\bibinfo {collaboration} {LUX}),\ }\href
  {https://doi.org/10.1103/PhysRevLett.118.021303} {\bibfield  {journal}
  {\bibinfo  {journal} {Phys. Rev. Lett.}\ }\textbf {\bibinfo {volume} {118}},\
  \bibinfo {pages} {021303} (\bibinfo {year} {2017})},\ \Eprint
  {https://arxiv.org/abs/1608.07648} {arXiv:1608.07648 [astro-ph.CO]}
  \BibitemShut {NoStop}%
\bibitem [{\citenamefont {Agnese}\ \emph {et~al.}(2018)\citenamefont {Agnese}
  \emph {et~al.}}]{SuperCDMS:2018mne}%
  \BibitemOpen
  \bibfield  {author} {\bibinfo {author} {\bibfnamefont {R.}~\bibnamefont
  {Agnese}} \emph {et~al.} (\bibinfo {collaboration} {SuperCDMS}),\ }\href
  {https://doi.org/10.1103/PhysRevLett.121.051301} {\bibfield  {journal}
  {\bibinfo  {journal} {Phys. Rev. Lett.}\ }\textbf {\bibinfo {volume} {121}},\
  \bibinfo {pages} {051301} (\bibinfo {year} {2018})},\ \bibinfo {note}
  {[Erratum: Phys.Rev.Lett. 122, 069901 (2019)]},\ \Eprint
  {https://arxiv.org/abs/1804.10697} {arXiv:1804.10697 [hep-ex]} \BibitemShut
  {NoStop}%
\bibitem [{\citenamefont {Aprile}\ \emph {et~al.}(2019)\citenamefont {Aprile}
  \emph {et~al.}}]{XENON:2019gfn}%
  \BibitemOpen
  \bibfield  {author} {\bibinfo {author} {\bibfnamefont {E.}~\bibnamefont
  {Aprile}} \emph {et~al.} (\bibinfo {collaboration} {XENON}),\ }\href
  {https://doi.org/10.1103/PhysRevLett.123.251801} {\bibfield  {journal}
  {\bibinfo  {journal} {Phys. Rev. Lett.}\ }\textbf {\bibinfo {volume} {123}},\
  \bibinfo {pages} {251801} (\bibinfo {year} {2019})},\ \Eprint
  {https://arxiv.org/abs/1907.11485} {arXiv:1907.11485 [hep-ex]} \BibitemShut
  {NoStop}%
\bibitem [{\citenamefont {Abdelhameed}\ \emph {et~al.}(2019)\citenamefont
  {Abdelhameed} \emph {et~al.}}]{CRESST:2019jnq}%
  \BibitemOpen
  \bibfield  {author} {\bibinfo {author} {\bibfnamefont {A.~H.}\ \bibnamefont
  {Abdelhameed}} \emph {et~al.} (\bibinfo {collaboration} {CRESST}),\ }\href
  {https://doi.org/10.1103/PhysRevD.100.102002} {\bibfield  {journal} {\bibinfo
   {journal} {Phys. Rev. D}\ }\textbf {\bibinfo {volume} {100}},\ \bibinfo
  {pages} {102002} (\bibinfo {year} {2019})},\ \Eprint
  {https://arxiv.org/abs/1904.00498} {arXiv:1904.00498 [astro-ph.CO]}
  \BibitemShut {NoStop}%
\bibitem [{\citenamefont {Wang}\ \emph {et~al.}(2020)\citenamefont {Wang} \emph
  {et~al.}}]{PandaX-II:2020oim}%
  \BibitemOpen
  \bibfield  {author} {\bibinfo {author} {\bibfnamefont {Q.}~\bibnamefont
  {Wang}} \emph {et~al.} (\bibinfo {collaboration} {PandaX-II}),\ }\href
  {https://doi.org/10.1088/1674-1137/abb658} {\bibfield  {journal} {\bibinfo
  {journal} {Chin. Phys. C}\ }\textbf {\bibinfo {volume} {44}},\ \bibinfo
  {pages} {125001} (\bibinfo {year} {2020})},\ \Eprint
  {https://arxiv.org/abs/2007.15469} {arXiv:2007.15469 [astro-ph.CO]}
  \BibitemShut {NoStop}%
\bibitem [{\citenamefont {Boehm}\ \emph {et~al.}(2004)\citenamefont {Boehm},
  \citenamefont {Fayet},\ and\ \citenamefont {Silk}}]{Boehm:2003ha}%
  \BibitemOpen
  \bibfield  {author} {\bibinfo {author} {\bibfnamefont {C.}~\bibnamefont
  {Boehm}}, \bibinfo {author} {\bibfnamefont {P.}~\bibnamefont {Fayet}},\ and\
  \bibinfo {author} {\bibfnamefont {J.}~\bibnamefont {Silk}},\ }\href
  {https://doi.org/10.1103/PhysRevD.69.101302} {\bibfield  {journal} {\bibinfo
  {journal} {Phys. Rev. D}\ }\textbf {\bibinfo {volume} {69}},\ \bibinfo
  {pages} {101302} (\bibinfo {year} {2004})},\ \Eprint
  {https://arxiv.org/abs/hep-ph/0311143} {arXiv:hep-ph/0311143} \BibitemShut
  {NoStop}%
\bibitem [{\citenamefont {Boehm}\ and\ \citenamefont
  {Fayet}(2004)}]{Boehm:2003hm}%
  \BibitemOpen
  \bibfield  {author} {\bibinfo {author} {\bibfnamefont {C.}~\bibnamefont
  {Boehm}}\ and\ \bibinfo {author} {\bibfnamefont {P.}~\bibnamefont {Fayet}},\
  }\href {https://doi.org/10.1016/j.nuclphysb.2004.01.015} {\bibfield
  {journal} {\bibinfo  {journal} {Nucl. Phys. B}\ }\textbf {\bibinfo {volume}
  {683}},\ \bibinfo {pages} {219} (\bibinfo {year} {2004})},\ \Eprint
  {https://arxiv.org/abs/hep-ph/0305261} {arXiv:hep-ph/0305261} \BibitemShut
  {NoStop}%
\bibitem [{\citenamefont {Strassler}\ and\ \citenamefont
  {Zurek}(2007)}]{Strassler:2006im}%
  \BibitemOpen
  \bibfield  {author} {\bibinfo {author} {\bibfnamefont {M.~J.}\ \bibnamefont
  {Strassler}}\ and\ \bibinfo {author} {\bibfnamefont {K.~M.}\ \bibnamefont
  {Zurek}},\ }\href {https://doi.org/10.1016/j.physletb.2007.06.055} {\bibfield
   {journal} {\bibinfo  {journal} {Phys. Lett. B}\ }\textbf {\bibinfo {volume}
  {651}},\ \bibinfo {pages} {374} (\bibinfo {year} {2007})},\ \Eprint
  {https://arxiv.org/abs/hep-ph/0604261} {arXiv:hep-ph/0604261} \BibitemShut
  {NoStop}%
\bibitem [{\citenamefont {Hooper}\ and\ \citenamefont
  {Zurek}(2008)}]{Hooper:2008im}%
  \BibitemOpen
  \bibfield  {author} {\bibinfo {author} {\bibfnamefont {D.}~\bibnamefont
  {Hooper}}\ and\ \bibinfo {author} {\bibfnamefont {K.~M.}\ \bibnamefont
  {Zurek}},\ }\href {https://doi.org/10.1103/PhysRevD.77.087302} {\bibfield
  {journal} {\bibinfo  {journal} {Phys. Rev. D}\ }\textbf {\bibinfo {volume}
  {77}},\ \bibinfo {pages} {087302} (\bibinfo {year} {2008})},\ \Eprint
  {https://arxiv.org/abs/0801.3686} {arXiv:0801.3686 [hep-ph]} \BibitemShut
  {NoStop}%
\bibitem [{\citenamefont {Feng}\ and\ \citenamefont
  {Kumar}(2008)}]{Feng:2008ya}%
  \BibitemOpen
  \bibfield  {author} {\bibinfo {author} {\bibfnamefont {J.~L.}\ \bibnamefont
  {Feng}}\ and\ \bibinfo {author} {\bibfnamefont {J.}~\bibnamefont {Kumar}},\
  }\href {https://doi.org/10.1103/PhysRevLett.101.231301} {\bibfield  {journal}
  {\bibinfo  {journal} {Phys. Rev. Lett.}\ }\textbf {\bibinfo {volume} {101}},\
  \bibinfo {pages} {231301} (\bibinfo {year} {2008})},\ \Eprint
  {https://arxiv.org/abs/0803.4196} {arXiv:0803.4196 [hep-ph]} \BibitemShut
  {NoStop}%
\bibitem [{\citenamefont {Falkowski}\ \emph {et~al.}(2011)\citenamefont
  {Falkowski}, \citenamefont {Ruderman},\ and\ \citenamefont
  {Volansky}}]{Falkowski:2011xh}%
  \BibitemOpen
  \bibfield  {author} {\bibinfo {author} {\bibfnamefont {A.}~\bibnamefont
  {Falkowski}}, \bibinfo {author} {\bibfnamefont {J.~T.}\ \bibnamefont
  {Ruderman}},\ and\ \bibinfo {author} {\bibfnamefont {T.}~\bibnamefont
  {Volansky}},\ }\href {https://doi.org/10.1007/JHEP05(2011)106} {\bibfield
  {journal} {\bibinfo  {journal} {JHEP}\ }\textbf {\bibinfo {volume} {05}},\
  \bibinfo {pages} {106}},\ \Eprint {https://arxiv.org/abs/1101.4936}
  {arXiv:1101.4936 [hep-ph]} \BibitemShut {NoStop}%
\bibitem [{\citenamefont {Lin}\ \emph {et~al.}(2012)\citenamefont {Lin},
  \citenamefont {Yu},\ and\ \citenamefont {Zurek}}]{Lin:2011gj}%
  \BibitemOpen
  \bibfield  {author} {\bibinfo {author} {\bibfnamefont {T.}~\bibnamefont
  {Lin}}, \bibinfo {author} {\bibfnamefont {H.-B.}\ \bibnamefont {Yu}},\ and\
  \bibinfo {author} {\bibfnamefont {K.~M.}\ \bibnamefont {Zurek}},\ }\href
  {https://doi.org/10.1103/PhysRevD.85.063503} {\bibfield  {journal} {\bibinfo
  {journal} {Phys. Rev. D}\ }\textbf {\bibinfo {volume} {85}},\ \bibinfo
  {pages} {063503} (\bibinfo {year} {2012})},\ \Eprint
  {https://arxiv.org/abs/1111.0293} {arXiv:1111.0293 [hep-ph]} \BibitemShut
  {NoStop}%
\bibitem [{\citenamefont {Hochberg}\ \emph {et~al.}(2014)\citenamefont
  {Hochberg}, \citenamefont {Kuflik}, \citenamefont {Volansky},\ and\
  \citenamefont {Wacker}}]{Hochberg:2014dra}%
  \BibitemOpen
  \bibfield  {author} {\bibinfo {author} {\bibfnamefont {Y.}~\bibnamefont
  {Hochberg}}, \bibinfo {author} {\bibfnamefont {E.}~\bibnamefont {Kuflik}},
  \bibinfo {author} {\bibfnamefont {T.}~\bibnamefont {Volansky}},\ and\
  \bibinfo {author} {\bibfnamefont {J.~G.}\ \bibnamefont {Wacker}},\ }\href
  {https://doi.org/10.1103/PhysRevLett.113.171301} {\bibfield  {journal}
  {\bibinfo  {journal} {Phys. Rev. Lett.}\ }\textbf {\bibinfo {volume} {113}},\
  \bibinfo {pages} {171301} (\bibinfo {year} {2014})},\ \Eprint
  {https://arxiv.org/abs/1402.5143} {arXiv:1402.5143 [hep-ph]} \BibitemShut
  {NoStop}%
\bibitem [{\citenamefont {D'Agnolo}\ and\ \citenamefont
  {Ruderman}(2015)}]{DAgnolo:2015ujb}%
  \BibitemOpen
  \bibfield  {author} {\bibinfo {author} {\bibfnamefont {R.~T.}\ \bibnamefont
  {D'Agnolo}}\ and\ \bibinfo {author} {\bibfnamefont {J.~T.}\ \bibnamefont
  {Ruderman}},\ }\href {https://doi.org/10.1103/PhysRevLett.115.061301}
  {\bibfield  {journal} {\bibinfo  {journal} {Phys. Rev. Lett.}\ }\textbf
  {\bibinfo {volume} {115}},\ \bibinfo {pages} {061301} (\bibinfo {year}
  {2015})},\ \Eprint {https://arxiv.org/abs/1505.07107} {arXiv:1505.07107
  [hep-ph]} \BibitemShut {NoStop}%
\bibitem [{\citenamefont {Kuflik}\ \emph {et~al.}(2016)\citenamefont {Kuflik},
  \citenamefont {Perelstein}, \citenamefont {Lorier},\ and\ \citenamefont
  {Tsai}}]{Kuflik:2015isi}%
  \BibitemOpen
  \bibfield  {author} {\bibinfo {author} {\bibfnamefont {E.}~\bibnamefont
  {Kuflik}}, \bibinfo {author} {\bibfnamefont {M.}~\bibnamefont {Perelstein}},
  \bibinfo {author} {\bibfnamefont {N.~R.-L.}\ \bibnamefont {Lorier}},\ and\
  \bibinfo {author} {\bibfnamefont {Y.-D.}\ \bibnamefont {Tsai}},\ }\href
  {https://doi.org/10.1103/PhysRevLett.116.221302} {\bibfield  {journal}
  {\bibinfo  {journal} {Phys. Rev. Lett.}\ }\textbf {\bibinfo {volume} {116}},\
  \bibinfo {pages} {221302} (\bibinfo {year} {2016})},\ \Eprint
  {https://arxiv.org/abs/1512.04545} {arXiv:1512.04545 [hep-ph]} \BibitemShut
  {NoStop}%
\bibitem [{\citenamefont {Green}\ and\ \citenamefont
  {Rajendran}(2017)}]{Green:2017ybv}%
  \BibitemOpen
  \bibfield  {author} {\bibinfo {author} {\bibfnamefont {D.}~\bibnamefont
  {Green}}\ and\ \bibinfo {author} {\bibfnamefont {S.}~\bibnamefont
  {Rajendran}},\ }\href {https://doi.org/10.1007/JHEP10(2017)013} {\bibfield
  {journal} {\bibinfo  {journal} {JHEP}\ }\textbf {\bibinfo {volume} {10}},\
  \bibinfo {pages} {013}},\ \Eprint {https://arxiv.org/abs/1701.08750}
  {arXiv:1701.08750 [hep-ph]} \BibitemShut {NoStop}%
\bibitem [{\citenamefont {D'Agnolo}\ \emph {et~al.}(2018)\citenamefont
  {D'Agnolo}, \citenamefont {Mondino}, \citenamefont {Ruderman},\ and\
  \citenamefont {Wang}}]{DAgnolo:2018wcn}%
  \BibitemOpen
  \bibfield  {author} {\bibinfo {author} {\bibfnamefont {R.~T.}\ \bibnamefont
  {D'Agnolo}}, \bibinfo {author} {\bibfnamefont {C.}~\bibnamefont {Mondino}},
  \bibinfo {author} {\bibfnamefont {J.~T.}\ \bibnamefont {Ruderman}},\ and\
  \bibinfo {author} {\bibfnamefont {P.-J.}\ \bibnamefont {Wang}},\ }\href
  {https://doi.org/10.1007/JHEP08(2018)079} {\bibfield  {journal} {\bibinfo
  {journal} {JHEP}\ }\textbf {\bibinfo {volume} {08}},\ \bibinfo {pages}
  {079}},\ \Eprint {https://arxiv.org/abs/1803.02901} {arXiv:1803.02901
  [hep-ph]} \BibitemShut {NoStop}%
\bibitem [{\citenamefont {Essig}\ \emph {et~al.}(2012)\citenamefont {Essig},
  \citenamefont {Mardon},\ and\ \citenamefont {Volansky}}]{Essig:2011nj}%
  \BibitemOpen
  \bibfield  {author} {\bibinfo {author} {\bibfnamefont {R.}~\bibnamefont
  {Essig}}, \bibinfo {author} {\bibfnamefont {J.}~\bibnamefont {Mardon}},\ and\
  \bibinfo {author} {\bibfnamefont {T.}~\bibnamefont {Volansky}},\ }\href
  {https://doi.org/10.1103/PhysRevD.85.076007} {\bibfield  {journal} {\bibinfo
  {journal} {Phys. Rev. D}\ }\textbf {\bibinfo {volume} {85}},\ \bibinfo
  {pages} {076007} (\bibinfo {year} {2012})},\ \Eprint
  {https://arxiv.org/abs/1108.5383} {arXiv:1108.5383 [hep-ph]} \BibitemShut
  {NoStop}%
\bibitem [{\citenamefont {Graham}\ \emph {et~al.}(2012)\citenamefont {Graham},
  \citenamefont {Kaplan}, \citenamefont {Rajendran},\ and\ \citenamefont
  {Walters}}]{Graham:2012su}%
  \BibitemOpen
  \bibfield  {author} {\bibinfo {author} {\bibfnamefont {P.~W.}\ \bibnamefont
  {Graham}}, \bibinfo {author} {\bibfnamefont {D.~E.}\ \bibnamefont {Kaplan}},
  \bibinfo {author} {\bibfnamefont {S.}~\bibnamefont {Rajendran}},\ and\
  \bibinfo {author} {\bibfnamefont {M.~T.}\ \bibnamefont {Walters}},\ }\href
  {https://doi.org/10.1016/j.dark.2012.09.001} {\bibfield  {journal} {\bibinfo
  {journal} {Phys. Dark Univ.}\ }\textbf {\bibinfo {volume} {1}},\ \bibinfo
  {pages} {32} (\bibinfo {year} {2012})},\ \Eprint
  {https://arxiv.org/abs/1203.2531} {arXiv:1203.2531 [hep-ph]} \BibitemShut
  {NoStop}%
\bibitem [{\citenamefont {Essig}\ \emph {et~al.}(2016)\citenamefont {Essig},
  \citenamefont {Fernandez-Serra}, \citenamefont {Mardon}, \citenamefont
  {Soto}, \citenamefont {Volansky},\ and\ \citenamefont {Yu}}]{Essig:2015cda}%
  \BibitemOpen
  \bibfield  {author} {\bibinfo {author} {\bibfnamefont {R.}~\bibnamefont
  {Essig}}, \bibinfo {author} {\bibfnamefont {M.}~\bibnamefont
  {Fernandez-Serra}}, \bibinfo {author} {\bibfnamefont {J.}~\bibnamefont
  {Mardon}}, \bibinfo {author} {\bibfnamefont {A.}~\bibnamefont {Soto}},
  \bibinfo {author} {\bibfnamefont {T.}~\bibnamefont {Volansky}},\ and\
  \bibinfo {author} {\bibfnamefont {T.-T.}\ \bibnamefont {Yu}},\ }\href
  {https://doi.org/10.1007/JHEP05(2016)046} {\bibfield  {journal} {\bibinfo
  {journal} {JHEP}\ }\textbf {\bibinfo {volume} {05}},\ \bibinfo {pages}
  {046}},\ \Eprint {https://arxiv.org/abs/1509.01598} {arXiv:1509.01598
  [hep-ph]} \BibitemShut {NoStop}%
\bibitem [{\citenamefont {Hochberg}\ \emph
  {et~al.}(2017{\natexlab{a}})\citenamefont {Hochberg}, \citenamefont {Lin},\
  and\ \citenamefont {Zurek}}]{Hochberg:2016sqx}%
  \BibitemOpen
  \bibfield  {author} {\bibinfo {author} {\bibfnamefont {Y.}~\bibnamefont
  {Hochberg}}, \bibinfo {author} {\bibfnamefont {T.}~\bibnamefont {Lin}},\ and\
  \bibinfo {author} {\bibfnamefont {K.~M.}\ \bibnamefont {Zurek}},\ }\href
  {https://doi.org/10.1103/PhysRevD.95.023013} {\bibfield  {journal} {\bibinfo
  {journal} {Phys. Rev. D}\ }\textbf {\bibinfo {volume} {95}},\ \bibinfo
  {pages} {023013} (\bibinfo {year} {2017}{\natexlab{a}})},\ \Eprint
  {https://arxiv.org/abs/1608.01994} {arXiv:1608.01994 [hep-ph]} \BibitemShut
  {NoStop}%
\bibitem [{\citenamefont {Bloch}\ \emph {et~al.}(2017)\citenamefont {Bloch},
  \citenamefont {Essig}, \citenamefont {Tobioka}, \citenamefont {Volansky},\
  and\ \citenamefont {Yu}}]{Bloch:2016sjj}%
  \BibitemOpen
  \bibfield  {author} {\bibinfo {author} {\bibfnamefont {I.~M.}\ \bibnamefont
  {Bloch}}, \bibinfo {author} {\bibfnamefont {R.}~\bibnamefont {Essig}},
  \bibinfo {author} {\bibfnamefont {K.}~\bibnamefont {Tobioka}}, \bibinfo
  {author} {\bibfnamefont {T.}~\bibnamefont {Volansky}},\ and\ \bibinfo
  {author} {\bibfnamefont {T.-T.}\ \bibnamefont {Yu}},\ }\href
  {https://doi.org/10.1007/JHEP06(2017)087} {\bibfield  {journal} {\bibinfo
  {journal} {JHEP}\ }\textbf {\bibinfo {volume} {06}},\ \bibinfo {pages}
  {087}},\ \Eprint {https://arxiv.org/abs/1608.02123} {arXiv:1608.02123
  [hep-ph]} \BibitemShut {NoStop}%
\bibitem [{\citenamefont {Knapen}\ \emph {et~al.}(2021)\citenamefont {Knapen},
  \citenamefont {Kozaczuk},\ and\ \citenamefont {Lin}}]{Knapen:2020aky}%
  \BibitemOpen
  \bibfield  {author} {\bibinfo {author} {\bibfnamefont {S.}~\bibnamefont
  {Knapen}}, \bibinfo {author} {\bibfnamefont {J.}~\bibnamefont {Kozaczuk}},\
  and\ \bibinfo {author} {\bibfnamefont {T.}~\bibnamefont {Lin}},\ }\href
  {https://doi.org/10.1103/PhysRevLett.127.081805} {\bibfield  {journal}
  {\bibinfo  {journal} {Phys. Rev. Lett.}\ }\textbf {\bibinfo {volume} {127}},\
  \bibinfo {pages} {081805} (\bibinfo {year} {2021})},\ \Eprint
  {https://arxiv.org/abs/2011.09496} {arXiv:2011.09496 [hep-ph]} \BibitemShut
  {NoStop}%
\bibitem [{\citenamefont {Liang}\ \emph {et~al.}(2022)\citenamefont {Liang},
  \citenamefont {Mo}, \citenamefont {Zheng},\ and\ \citenamefont
  {Zhang}}]{Liang:2022xbu}%
  \BibitemOpen
  \bibfield  {author} {\bibinfo {author} {\bibfnamefont {Z.-L.}\ \bibnamefont
  {Liang}}, \bibinfo {author} {\bibfnamefont {C.}~\bibnamefont {Mo}}, \bibinfo
  {author} {\bibfnamefont {F.}~\bibnamefont {Zheng}},\ and\ \bibinfo {author}
  {\bibfnamefont {P.}~\bibnamefont {Zhang}},\ }\href
  {https://doi.org/10.1103/PhysRevD.106.043004} {\bibfield  {journal} {\bibinfo
   {journal} {Phys. Rev. D}\ }\textbf {\bibinfo {volume} {106}},\ \bibinfo
  {pages} {043004} (\bibinfo {year} {2022})},\ \Eprint
  {https://arxiv.org/abs/2205.03395} {arXiv:2205.03395 [hep-ph]} \BibitemShut
  {NoStop}%
\bibitem [{\citenamefont {Berghaus}\ \emph {et~al.}(2022)\citenamefont
  {Berghaus}, \citenamefont {Esposito}, \citenamefont {Essig},\ and\
  \citenamefont {Sholapurkar}}]{Berghaus:2022pbu}%
  \BibitemOpen
  \bibfield  {author} {\bibinfo {author} {\bibfnamefont {K.~V.}\ \bibnamefont
  {Berghaus}}, \bibinfo {author} {\bibfnamefont {A.}~\bibnamefont {Esposito}},
  \bibinfo {author} {\bibfnamefont {R.}~\bibnamefont {Essig}},\ and\ \bibinfo
  {author} {\bibfnamefont {M.}~\bibnamefont {Sholapurkar}},\ }\href@noop {} {\
  (\bibinfo {year} {2022})},\ \Eprint {https://arxiv.org/abs/2210.06490}
  {arXiv:2210.06490 [hep-ph]} \BibitemShut {NoStop}%
\bibitem [{\citenamefont {Hochberg}\ \emph
  {et~al.}(2016{\natexlab{a}})\citenamefont {Hochberg}, \citenamefont {Zhao},\
  and\ \citenamefont {Zurek}}]{Hochberg:2015pha}%
  \BibitemOpen
  \bibfield  {author} {\bibinfo {author} {\bibfnamefont {Y.}~\bibnamefont
  {Hochberg}}, \bibinfo {author} {\bibfnamefont {Y.}~\bibnamefont {Zhao}},\
  and\ \bibinfo {author} {\bibfnamefont {K.~M.}\ \bibnamefont {Zurek}},\ }\href
  {https://doi.org/10.1103/PhysRevLett.116.011301} {\bibfield  {journal}
  {\bibinfo  {journal} {Phys. Rev. Lett.}\ }\textbf {\bibinfo {volume} {116}},\
  \bibinfo {pages} {011301} (\bibinfo {year} {2016}{\natexlab{a}})},\ \Eprint
  {https://arxiv.org/abs/1504.07237} {arXiv:1504.07237 [hep-ph]} \BibitemShut
  {NoStop}%
\bibitem [{\citenamefont {Hochberg}\ \emph
  {et~al.}(2016{\natexlab{b}})\citenamefont {Hochberg}, \citenamefont {Lin},\
  and\ \citenamefont {Zurek}}]{Hochberg:2016ajh}%
  \BibitemOpen
  \bibfield  {author} {\bibinfo {author} {\bibfnamefont {Y.}~\bibnamefont
  {Hochberg}}, \bibinfo {author} {\bibfnamefont {T.}~\bibnamefont {Lin}},\ and\
  \bibinfo {author} {\bibfnamefont {K.~M.}\ \bibnamefont {Zurek}},\ }\href
  {https://doi.org/10.1103/PhysRevD.94.015019} {\bibfield  {journal} {\bibinfo
  {journal} {Phys. Rev. D}\ }\textbf {\bibinfo {volume} {94}},\ \bibinfo
  {pages} {015019} (\bibinfo {year} {2016}{\natexlab{b}})},\ \Eprint
  {https://arxiv.org/abs/1604.06800} {arXiv:1604.06800 [hep-ph]} \BibitemShut
  {NoStop}%
\bibitem [{\citenamefont {Hochberg}\ \emph {et~al.}(2019)\citenamefont
  {Hochberg}, \citenamefont {Charaev}, \citenamefont {Nam}, \citenamefont
  {Verma}, \citenamefont {Colangelo},\ and\ \citenamefont
  {Berggren}}]{Hochberg:2019cyy}%
  \BibitemOpen
  \bibfield  {author} {\bibinfo {author} {\bibfnamefont {Y.}~\bibnamefont
  {Hochberg}}, \bibinfo {author} {\bibfnamefont {I.}~\bibnamefont {Charaev}},
  \bibinfo {author} {\bibfnamefont {S.-W.}\ \bibnamefont {Nam}}, \bibinfo
  {author} {\bibfnamefont {V.}~\bibnamefont {Verma}}, \bibinfo {author}
  {\bibfnamefont {M.}~\bibnamefont {Colangelo}},\ and\ \bibinfo {author}
  {\bibfnamefont {K.~K.}\ \bibnamefont {Berggren}},\ }\href
  {https://doi.org/10.1103/PhysRevLett.123.151802} {\bibfield  {journal}
  {\bibinfo  {journal} {Phys. Rev. Lett.}\ }\textbf {\bibinfo {volume} {123}},\
  \bibinfo {pages} {151802} (\bibinfo {year} {2019})},\ \Eprint
  {https://arxiv.org/abs/1903.05101} {arXiv:1903.05101 [hep-ph]} \BibitemShut
  {NoStop}%
\bibitem [{\citenamefont {Griffin}\ \emph {et~al.}(2021)\citenamefont
  {Griffin}, \citenamefont {Hochberg}, \citenamefont {Inzani}, \citenamefont
  {Kurinsky}, \citenamefont {Lin},\ and\ \citenamefont
  {Chin}}]{Griffin:2020lgd}%
  \BibitemOpen
  \bibfield  {author} {\bibinfo {author} {\bibfnamefont {S.~M.}\ \bibnamefont
  {Griffin}}, \bibinfo {author} {\bibfnamefont {Y.}~\bibnamefont {Hochberg}},
  \bibinfo {author} {\bibfnamefont {K.}~\bibnamefont {Inzani}}, \bibinfo
  {author} {\bibfnamefont {N.}~\bibnamefont {Kurinsky}}, \bibinfo {author}
  {\bibfnamefont {T.}~\bibnamefont {Lin}},\ and\ \bibinfo {author}
  {\bibfnamefont {T.}~\bibnamefont {Chin}},\ }\href
  {https://doi.org/10.1103/PhysRevD.103.075002} {\bibfield  {journal} {\bibinfo
   {journal} {Phys. Rev. D}\ }\textbf {\bibinfo {volume} {103}},\ \bibinfo
  {pages} {075002} (\bibinfo {year} {2021})},\ \Eprint
  {https://arxiv.org/abs/2008.08560} {arXiv:2008.08560 [hep-ph]} \BibitemShut
  {NoStop}%
\bibitem [{\citenamefont {Hochberg}\ \emph {et~al.}(2021)\citenamefont
  {Hochberg}, \citenamefont {Lehmann}, \citenamefont {Charaev}, \citenamefont
  {Chiles}, \citenamefont {Colangelo}, \citenamefont {Nam},\ and\ \citenamefont
  {Berggren}}]{Hochberg:2021yud}%
  \BibitemOpen
  \bibfield  {author} {\bibinfo {author} {\bibfnamefont {Y.}~\bibnamefont
  {Hochberg}}, \bibinfo {author} {\bibfnamefont {B.~V.}\ \bibnamefont
  {Lehmann}}, \bibinfo {author} {\bibfnamefont {I.}~\bibnamefont {Charaev}},
  \bibinfo {author} {\bibfnamefont {J.}~\bibnamefont {Chiles}}, \bibinfo
  {author} {\bibfnamefont {M.}~\bibnamefont {Colangelo}}, \bibinfo {author}
  {\bibfnamefont {S.~W.}\ \bibnamefont {Nam}},\ and\ \bibinfo {author}
  {\bibfnamefont {K.~K.}\ \bibnamefont {Berggren}},\ }\href@noop {} {\
  (\bibinfo {year} {2021})},\ \Eprint {https://arxiv.org/abs/2110.01586}
  {arXiv:2110.01586 [hep-ph]} \BibitemShut {NoStop}%
\bibitem [{\citenamefont {Hochberg}\ \emph {et~al.}(2018)\citenamefont
  {Hochberg}, \citenamefont {Kahn}, \citenamefont {Lisanti}, \citenamefont
  {Zurek}, \citenamefont {Grushin}, \citenamefont {Ilan}, \citenamefont
  {Griffin}, \citenamefont {Liu}, \citenamefont {Weber},\ and\ \citenamefont
  {Neaton}}]{Hochberg:2017wce}%
  \BibitemOpen
  \bibfield  {author} {\bibinfo {author} {\bibfnamefont {Y.}~\bibnamefont
  {Hochberg}}, \bibinfo {author} {\bibfnamefont {Y.}~\bibnamefont {Kahn}},
  \bibinfo {author} {\bibfnamefont {M.}~\bibnamefont {Lisanti}}, \bibinfo
  {author} {\bibfnamefont {K.~M.}\ \bibnamefont {Zurek}}, \bibinfo {author}
  {\bibfnamefont {A.~G.}\ \bibnamefont {Grushin}}, \bibinfo {author}
  {\bibfnamefont {R.}~\bibnamefont {Ilan}}, \bibinfo {author} {\bibfnamefont
  {S.~M.}\ \bibnamefont {Griffin}}, \bibinfo {author} {\bibfnamefont {Z.-F.}\
  \bibnamefont {Liu}}, \bibinfo {author} {\bibfnamefont {S.~F.}\ \bibnamefont
  {Weber}},\ and\ \bibinfo {author} {\bibfnamefont {J.~B.}\ \bibnamefont
  {Neaton}},\ }\href {https://doi.org/10.1103/PhysRevD.97.015004} {\bibfield
  {journal} {\bibinfo  {journal} {Phys. Rev. D}\ }\textbf {\bibinfo {volume}
  {97}},\ \bibinfo {pages} {015004} (\bibinfo {year} {2018})},\ \Eprint
  {https://arxiv.org/abs/1708.08929} {arXiv:1708.08929 [hep-ph]} \BibitemShut
  {NoStop}%
\bibitem [{\citenamefont {Coskuner}\ \emph {et~al.}(2021)\citenamefont
  {Coskuner}, \citenamefont {Mitridate}, \citenamefont {Olivares},\ and\
  \citenamefont {Zurek}}]{Coskuner:2019odd}%
  \BibitemOpen
  \bibfield  {author} {\bibinfo {author} {\bibfnamefont {A.}~\bibnamefont
  {Coskuner}}, \bibinfo {author} {\bibfnamefont {A.}~\bibnamefont {Mitridate}},
  \bibinfo {author} {\bibfnamefont {A.}~\bibnamefont {Olivares}},\ and\
  \bibinfo {author} {\bibfnamefont {K.~M.}\ \bibnamefont {Zurek}},\ }\href
  {https://doi.org/10.1103/PhysRevD.103.016006} {\bibfield  {journal} {\bibinfo
   {journal} {Phys. Rev. D}\ }\textbf {\bibinfo {volume} {103}},\ \bibinfo
  {pages} {016006} (\bibinfo {year} {2021})},\ \Eprint
  {https://arxiv.org/abs/1909.09170} {arXiv:1909.09170 [hep-ph]} \BibitemShut
  {NoStop}%
\bibitem [{\citenamefont {Geilhufe}\ \emph {et~al.}(2020)\citenamefont
  {Geilhufe}, \citenamefont {Kahlhoefer},\ and\ \citenamefont
  {Winkler}}]{Geilhufe:2019ndy}%
  \BibitemOpen
  \bibfield  {author} {\bibinfo {author} {\bibfnamefont {R.~M.}\ \bibnamefont
  {Geilhufe}}, \bibinfo {author} {\bibfnamefont {F.}~\bibnamefont
  {Kahlhoefer}},\ and\ \bibinfo {author} {\bibfnamefont {M.~W.}\ \bibnamefont
  {Winkler}},\ }\href {https://doi.org/10.1103/PhysRevD.101.055005} {\bibfield
  {journal} {\bibinfo  {journal} {Phys. Rev. D}\ }\textbf {\bibinfo {volume}
  {101}},\ \bibinfo {pages} {055005} (\bibinfo {year} {2020})},\ \Eprint
  {https://arxiv.org/abs/1910.02091} {arXiv:1910.02091 [hep-ph]} \BibitemShut
  {NoStop}%
\bibitem [{\citenamefont {Capparelli}\ \emph {et~al.}(2015)\citenamefont
  {Capparelli}, \citenamefont {Cavoto}, \citenamefont {Mazzilli},\ and\
  \citenamefont {Polosa}}]{Capparelli:2014lua}%
  \BibitemOpen
  \bibfield  {author} {\bibinfo {author} {\bibfnamefont {L.~M.}\ \bibnamefont
  {Capparelli}}, \bibinfo {author} {\bibfnamefont {G.}~\bibnamefont {Cavoto}},
  \bibinfo {author} {\bibfnamefont {D.}~\bibnamefont {Mazzilli}},\ and\
  \bibinfo {author} {\bibfnamefont {A.~D.}\ \bibnamefont {Polosa}},\ }\href
  {https://doi.org/10.1016/j.dark.2015.08.002} {\bibfield  {journal} {\bibinfo
  {journal} {Phys. Dark Univ.}\ }\textbf {\bibinfo {volume} {9-10}},\ \bibinfo
  {pages} {24} (\bibinfo {year} {2015})},\ \bibinfo {note} {[Erratum: Phys.Dark
  Univ. 11, 79--80 (2016)]},\ \Eprint {https://arxiv.org/abs/1412.8213}
  {arXiv:1412.8213 [physics.ins-det]} \BibitemShut {NoStop}%
\bibitem [{\citenamefont {Hochberg}\ \emph
  {et~al.}(2017{\natexlab{b}})\citenamefont {Hochberg}, \citenamefont {Kahn},
  \citenamefont {Lisanti}, \citenamefont {Tully},\ and\ \citenamefont
  {Zurek}}]{Hochberg:2016ntt}%
  \BibitemOpen
  \bibfield  {author} {\bibinfo {author} {\bibfnamefont {Y.}~\bibnamefont
  {Hochberg}}, \bibinfo {author} {\bibfnamefont {Y.}~\bibnamefont {Kahn}},
  \bibinfo {author} {\bibfnamefont {M.}~\bibnamefont {Lisanti}}, \bibinfo
  {author} {\bibfnamefont {C.~G.}\ \bibnamefont {Tully}},\ and\ \bibinfo
  {author} {\bibfnamefont {K.~M.}\ \bibnamefont {Zurek}},\ }\href
  {https://doi.org/10.1016/j.physletb.2017.06.051} {\bibfield  {journal}
  {\bibinfo  {journal} {Phys. Lett. B}\ }\textbf {\bibinfo {volume} {772}},\
  \bibinfo {pages} {239} (\bibinfo {year} {2017}{\natexlab{b}})},\ \Eprint
  {https://arxiv.org/abs/1606.08849} {arXiv:1606.08849 [hep-ph]} \BibitemShut
  {NoStop}%
\bibitem [{\citenamefont {Cavoto}\ \emph {et~al.}(2016)\citenamefont {Cavoto},
  \citenamefont {Cirillo}, \citenamefont {Cocina}, \citenamefont {Ferretti},\
  and\ \citenamefont {Polosa}}]{Cavoto:2016lqo}%
  \BibitemOpen
  \bibfield  {author} {\bibinfo {author} {\bibfnamefont {G.}~\bibnamefont
  {Cavoto}}, \bibinfo {author} {\bibfnamefont {E.~N.~M.}\ \bibnamefont
  {Cirillo}}, \bibinfo {author} {\bibfnamefont {F.}~\bibnamefont {Cocina}},
  \bibinfo {author} {\bibfnamefont {J.}~\bibnamefont {Ferretti}},\ and\
  \bibinfo {author} {\bibfnamefont {A.~D.}\ \bibnamefont {Polosa}},\ }\href
  {https://doi.org/10.1140/epjc/s10052-016-4193-7} {\bibfield  {journal}
  {\bibinfo  {journal} {Eur. Phys. J. C}\ }\textbf {\bibinfo {volume} {76}},\
  \bibinfo {pages} {349} (\bibinfo {year} {2016})},\ \Eprint
  {https://arxiv.org/abs/1602.03216} {arXiv:1602.03216 [physics.ins-det]}
  \BibitemShut {NoStop}%
\bibitem [{\citenamefont {Cavoto}\ \emph {et~al.}(2018)\citenamefont {Cavoto},
  \citenamefont {Luchetta},\ and\ \citenamefont {Polosa}}]{Cavoto:2017otc}%
  \BibitemOpen
  \bibfield  {author} {\bibinfo {author} {\bibfnamefont {G.}~\bibnamefont
  {Cavoto}}, \bibinfo {author} {\bibfnamefont {F.}~\bibnamefont {Luchetta}},\
  and\ \bibinfo {author} {\bibfnamefont {A.~D.}\ \bibnamefont {Polosa}},\
  }\href {https://doi.org/10.1016/j.physletb.2017.11.064} {\bibfield  {journal}
  {\bibinfo  {journal} {Phys. Lett. B}\ }\textbf {\bibinfo {volume} {776}},\
  \bibinfo {pages} {338} (\bibinfo {year} {2018})},\ \Eprint
  {https://arxiv.org/abs/1706.02487} {arXiv:1706.02487 [hep-ph]} \BibitemShut
  {NoStop}%
\bibitem [{\citenamefont {Arvanitaki}\ \emph {et~al.}(2018)\citenamefont
  {Arvanitaki}, \citenamefont {Dimopoulos},\ and\ \citenamefont
  {Van~Tilburg}}]{Arvanitaki:2017nhi}%
  \BibitemOpen
  \bibfield  {author} {\bibinfo {author} {\bibfnamefont {A.}~\bibnamefont
  {Arvanitaki}}, \bibinfo {author} {\bibfnamefont {S.}~\bibnamefont
  {Dimopoulos}},\ and\ \bibinfo {author} {\bibfnamefont {K.}~\bibnamefont
  {Van~Tilburg}},\ }\href {https://doi.org/10.1103/PhysRevX.8.041001}
  {\bibfield  {journal} {\bibinfo  {journal} {Phys. Rev. X}\ }\textbf {\bibinfo
  {volume} {8}},\ \bibinfo {pages} {041001} (\bibinfo {year} {2018})},\ \Eprint
  {https://arxiv.org/abs/1709.05354} {arXiv:1709.05354 [hep-ph]} \BibitemShut
  {NoStop}%
\bibitem [{\citenamefont {Bunting}\ \emph {et~al.}(2017)\citenamefont
  {Bunting}, \citenamefont {Gratta}, \citenamefont {Melia},\ and\ \citenamefont
  {Rajendran}}]{Bunting:2017net}%
  \BibitemOpen
  \bibfield  {author} {\bibinfo {author} {\bibfnamefont {P.~C.}\ \bibnamefont
  {Bunting}}, \bibinfo {author} {\bibfnamefont {G.}~\bibnamefont {Gratta}},
  \bibinfo {author} {\bibfnamefont {T.}~\bibnamefont {Melia}},\ and\ \bibinfo
  {author} {\bibfnamefont {S.}~\bibnamefont {Rajendran}},\ }\href
  {https://doi.org/10.1103/PhysRevD.95.095001} {\bibfield  {journal} {\bibinfo
  {journal} {Phys. Rev. D}\ }\textbf {\bibinfo {volume} {95}},\ \bibinfo
  {pages} {095001} (\bibinfo {year} {2017})},\ \Eprint
  {https://arxiv.org/abs/1701.06566} {arXiv:1701.06566 [hep-ph]} \BibitemShut
  {NoStop}%
\bibitem [{\citenamefont {Chen}\ \emph {et~al.}(2020)\citenamefont {Chen},
  \citenamefont {Mahapatra}, \citenamefont {Agnolet}, \citenamefont {Nippe},
  \citenamefont {Lu}, \citenamefont {Bunting}, \citenamefont {Melia},
  \citenamefont {Rajendran}, \citenamefont {Gratta},\ and\ \citenamefont
  {Long}}]{Chen:2020jia}%
  \BibitemOpen
  \bibfield  {author} {\bibinfo {author} {\bibfnamefont {H.}~\bibnamefont
  {Chen}}, \bibinfo {author} {\bibfnamefont {R.}~\bibnamefont {Mahapatra}},
  \bibinfo {author} {\bibfnamefont {G.}~\bibnamefont {Agnolet}}, \bibinfo
  {author} {\bibfnamefont {M.}~\bibnamefont {Nippe}}, \bibinfo {author}
  {\bibfnamefont {M.}~\bibnamefont {Lu}}, \bibinfo {author} {\bibfnamefont
  {P.~C.}\ \bibnamefont {Bunting}}, \bibinfo {author} {\bibfnamefont
  {T.}~\bibnamefont {Melia}}, \bibinfo {author} {\bibfnamefont
  {S.}~\bibnamefont {Rajendran}}, \bibinfo {author} {\bibfnamefont
  {G.}~\bibnamefont {Gratta}},\ and\ \bibinfo {author} {\bibfnamefont
  {J.}~\bibnamefont {Long}},\ }\href@noop {} {\  (\bibinfo {year} {2020})},\
  \Eprint {https://arxiv.org/abs/2002.09409} {arXiv:2002.09409
  [physics.ins-det]} \BibitemShut {NoStop}%
\bibitem [{\citenamefont {Guo}\ and\ \citenamefont
  {McKinsey}(2013)}]{Guo:2013dt}%
  \BibitemOpen
  \bibfield  {author} {\bibinfo {author} {\bibfnamefont {W.}~\bibnamefont
  {Guo}}\ and\ \bibinfo {author} {\bibfnamefont {D.~N.}\ \bibnamefont
  {McKinsey}},\ }\href {https://doi.org/10.1103/PhysRevD.87.115001} {\bibfield
  {journal} {\bibinfo  {journal} {Phys. Rev. D}\ }\textbf {\bibinfo {volume}
  {87}},\ \bibinfo {pages} {115001} (\bibinfo {year} {2013})},\ \Eprint
  {https://arxiv.org/abs/1302.0534} {arXiv:1302.0534 [astro-ph.IM]}
  \BibitemShut {NoStop}%
\bibitem [{\citenamefont {Schutz}\ and\ \citenamefont
  {Zurek}(2016)}]{Schutz:2016tid}%
  \BibitemOpen
  \bibfield  {author} {\bibinfo {author} {\bibfnamefont {K.}~\bibnamefont
  {Schutz}}\ and\ \bibinfo {author} {\bibfnamefont {K.~M.}\ \bibnamefont
  {Zurek}},\ }\href {https://doi.org/10.1103/PhysRevLett.117.121302} {\bibfield
   {journal} {\bibinfo  {journal} {Phys. Rev. Lett.}\ }\textbf {\bibinfo
  {volume} {117}},\ \bibinfo {pages} {121302} (\bibinfo {year} {2016})},\
  \Eprint {https://arxiv.org/abs/1604.08206} {arXiv:1604.08206 [hep-ph]}
  \BibitemShut {NoStop}%
\bibitem [{\citenamefont {Knapen}\ \emph {et~al.}(2017)\citenamefont {Knapen},
  \citenamefont {Lin},\ and\ \citenamefont {Zurek}}]{Knapen:2016cue}%
  \BibitemOpen
  \bibfield  {author} {\bibinfo {author} {\bibfnamefont {S.}~\bibnamefont
  {Knapen}}, \bibinfo {author} {\bibfnamefont {T.}~\bibnamefont {Lin}},\ and\
  \bibinfo {author} {\bibfnamefont {K.~M.}\ \bibnamefont {Zurek}},\ }\href
  {https://doi.org/10.1103/PhysRevD.95.056019} {\bibfield  {journal} {\bibinfo
  {journal} {Phys. Rev. D}\ }\textbf {\bibinfo {volume} {95}},\ \bibinfo
  {pages} {056019} (\bibinfo {year} {2017})},\ \Eprint
  {https://arxiv.org/abs/1611.06228} {arXiv:1611.06228 [hep-ph]} \BibitemShut
  {NoStop}%
\bibitem [{\citenamefont {Hertel}\ \emph {et~al.}(2019)\citenamefont {Hertel},
  \citenamefont {Biekert}, \citenamefont {Lin}, \citenamefont {Velan},\ and\
  \citenamefont {McKinsey}}]{Hertel:2018aal}%
  \BibitemOpen
  \bibfield  {author} {\bibinfo {author} {\bibfnamefont {S.~A.}\ \bibnamefont
  {Hertel}}, \bibinfo {author} {\bibfnamefont {A.}~\bibnamefont {Biekert}},
  \bibinfo {author} {\bibfnamefont {J.}~\bibnamefont {Lin}}, \bibinfo {author}
  {\bibfnamefont {V.}~\bibnamefont {Velan}},\ and\ \bibinfo {author}
  {\bibfnamefont {D.~N.}\ \bibnamefont {McKinsey}},\ }\href
  {https://doi.org/10.1103/PhysRevD.100.092007} {\bibfield  {journal} {\bibinfo
   {journal} {Phys. Rev. D}\ }\textbf {\bibinfo {volume} {100}},\ \bibinfo
  {pages} {092007} (\bibinfo {year} {2019})},\ \Eprint
  {https://arxiv.org/abs/1810.06283} {arXiv:1810.06283 [physics.ins-det]}
  \BibitemShut {NoStop}%
\bibitem [{\citenamefont {Acanfora}\ \emph {et~al.}(2019)\citenamefont
  {Acanfora}, \citenamefont {Esposito},\ and\ \citenamefont
  {Polosa}}]{Acanfora:2019con}%
  \BibitemOpen
  \bibfield  {author} {\bibinfo {author} {\bibfnamefont {F.}~\bibnamefont
  {Acanfora}}, \bibinfo {author} {\bibfnamefont {A.}~\bibnamefont {Esposito}},\
  and\ \bibinfo {author} {\bibfnamefont {A.~D.}\ \bibnamefont {Polosa}},\
  }\href {https://doi.org/10.1140/epjc/s10052-019-7057-0} {\bibfield  {journal}
  {\bibinfo  {journal} {Eur. Phys. J. C}\ }\textbf {\bibinfo {volume} {79}},\
  \bibinfo {pages} {549} (\bibinfo {year} {2019})},\ \Eprint
  {https://arxiv.org/abs/1902.02361} {arXiv:1902.02361 [hep-ph]} \BibitemShut
  {NoStop}%
\bibitem [{\citenamefont {Caputo}\ \emph {et~al.}(2019)\citenamefont {Caputo},
  \citenamefont {Esposito},\ and\ \citenamefont {Polosa}}]{Caputo:2019cyg}%
  \BibitemOpen
  \bibfield  {author} {\bibinfo {author} {\bibfnamefont {A.}~\bibnamefont
  {Caputo}}, \bibinfo {author} {\bibfnamefont {A.}~\bibnamefont {Esposito}},\
  and\ \bibinfo {author} {\bibfnamefont {A.~D.}\ \bibnamefont {Polosa}},\
  }\href {https://doi.org/10.1103/PhysRevD.100.116007} {\bibfield  {journal}
  {\bibinfo  {journal} {Phys. Rev. D}\ }\textbf {\bibinfo {volume} {100}},\
  \bibinfo {pages} {116007} (\bibinfo {year} {2019})},\ \Eprint
  {https://arxiv.org/abs/1907.10635} {arXiv:1907.10635 [hep-ph]} \BibitemShut
  {NoStop}%
\bibitem [{\citenamefont {Caputo}\ \emph {et~al.}(2020)\citenamefont {Caputo},
  \citenamefont {Esposito}, \citenamefont {Geoffray}, \citenamefont {Polosa},\
  and\ \citenamefont {Sun}}]{Caputo:2019xum}%
  \BibitemOpen
  \bibfield  {author} {\bibinfo {author} {\bibfnamefont {A.}~\bibnamefont
  {Caputo}}, \bibinfo {author} {\bibfnamefont {A.}~\bibnamefont {Esposito}},
  \bibinfo {author} {\bibfnamefont {E.}~\bibnamefont {Geoffray}}, \bibinfo
  {author} {\bibfnamefont {A.~D.}\ \bibnamefont {Polosa}},\ and\ \bibinfo
  {author} {\bibfnamefont {S.}~\bibnamefont {Sun}},\ }\href
  {https://doi.org/10.1016/j.physletb.2020.135258} {\bibfield  {journal}
  {\bibinfo  {journal} {Phys. Lett. B}\ }\textbf {\bibinfo {volume} {802}},\
  \bibinfo {pages} {135258} (\bibinfo {year} {2020})},\ \Eprint
  {https://arxiv.org/abs/1911.04511} {arXiv:1911.04511 [hep-ph]} \BibitemShut
  {NoStop}%
\bibitem [{\citenamefont {Baym}\ \emph {et~al.}(2020)\citenamefont {Baym},
  \citenamefont {Beck}, \citenamefont {Filippini}, \citenamefont {Pethick},\
  and\ \citenamefont {Shelton}}]{Baym:2020uos}%
  \BibitemOpen
  \bibfield  {author} {\bibinfo {author} {\bibfnamefont {G.}~\bibnamefont
  {Baym}}, \bibinfo {author} {\bibfnamefont {D.~H.}\ \bibnamefont {Beck}},
  \bibinfo {author} {\bibfnamefont {J.~P.}\ \bibnamefont {Filippini}}, \bibinfo
  {author} {\bibfnamefont {C.~J.}\ \bibnamefont {Pethick}},\ and\ \bibinfo
  {author} {\bibfnamefont {J.}~\bibnamefont {Shelton}},\ }\href
  {https://doi.org/10.1103/PhysRevD.104.019901} {\bibfield  {journal} {\bibinfo
   {journal} {Phys. Rev. D}\ }\textbf {\bibinfo {volume} {102}},\ \bibinfo
  {pages} {035014} (\bibinfo {year} {2020})},\ \bibinfo {note} {[Erratum:
  Phys.Rev.D 104, 019901 (2021)]},\ \Eprint {https://arxiv.org/abs/2005.08824}
  {arXiv:2005.08824 [hep-ph]} \BibitemShut {NoStop}%
\bibitem [{\citenamefont {Caputo}\ \emph {et~al.}(2021)\citenamefont {Caputo},
  \citenamefont {Esposito}, \citenamefont {Piccinini}, \citenamefont {Polosa},\
  and\ \citenamefont {Rossi}}]{Caputo:2020sys}%
  \BibitemOpen
  \bibfield  {author} {\bibinfo {author} {\bibfnamefont {A.}~\bibnamefont
  {Caputo}}, \bibinfo {author} {\bibfnamefont {A.}~\bibnamefont {Esposito}},
  \bibinfo {author} {\bibfnamefont {F.}~\bibnamefont {Piccinini}}, \bibinfo
  {author} {\bibfnamefont {A.~D.}\ \bibnamefont {Polosa}},\ and\ \bibinfo
  {author} {\bibfnamefont {G.}~\bibnamefont {Rossi}},\ }\href
  {https://doi.org/10.1103/PhysRevD.103.055017} {\bibfield  {journal} {\bibinfo
   {journal} {Phys. Rev. D}\ }\textbf {\bibinfo {volume} {103}},\ \bibinfo
  {pages} {055017} (\bibinfo {year} {2021})},\ \Eprint
  {https://arxiv.org/abs/2012.01432} {arXiv:2012.01432 [hep-ph]} \BibitemShut
  {NoStop}%
\bibitem [{\citenamefont {Matchev}\ \emph {et~al.}(2022)\citenamefont
  {Matchev}, \citenamefont {Smolinsky}, \citenamefont {Xue},\ and\
  \citenamefont {You}}]{Matchev:2021fuw}%
  \BibitemOpen
  \bibfield  {author} {\bibinfo {author} {\bibfnamefont {K.~T.}\ \bibnamefont
  {Matchev}}, \bibinfo {author} {\bibfnamefont {J.}~\bibnamefont {Smolinsky}},
  \bibinfo {author} {\bibfnamefont {W.}~\bibnamefont {Xue}},\ and\ \bibinfo
  {author} {\bibfnamefont {Y.}~\bibnamefont {You}},\ }\href
  {https://doi.org/10.1007/JHEP05(2022)034} {\bibfield  {journal} {\bibinfo
  {journal} {JHEP}\ }\textbf {\bibinfo {volume} {05}},\ \bibinfo {pages}
  {034}},\ \Eprint {https://arxiv.org/abs/2108.07275} {arXiv:2108.07275
  [hep-ph]} \BibitemShut {NoStop}%
\bibitem [{\citenamefont {You}\ \emph {et~al.}(2022)\citenamefont {You},
  \citenamefont {Smolinsky}, \citenamefont {Xue}, \citenamefont {Matchev},
  \citenamefont {Gunther}, \citenamefont {Lee},\ and\ \citenamefont
  {Saab}}]{You:2022pyn}%
  \BibitemOpen
  \bibfield  {author} {\bibinfo {author} {\bibfnamefont {Y.}~\bibnamefont
  {You}}, \bibinfo {author} {\bibfnamefont {J.}~\bibnamefont {Smolinsky}},
  \bibinfo {author} {\bibfnamefont {W.}~\bibnamefont {Xue}}, \bibinfo {author}
  {\bibfnamefont {K.~T.}\ \bibnamefont {Matchev}}, \bibinfo {author}
  {\bibfnamefont {K.}~\bibnamefont {Gunther}}, \bibinfo {author} {\bibfnamefont
  {Y.}~\bibnamefont {Lee}},\ and\ \bibinfo {author} {\bibfnamefont
  {T.}~\bibnamefont {Saab}},\ }\href@noop {} {\  (\bibinfo {year} {2022})},\
  \Eprint {https://arxiv.org/abs/2208.14474} {arXiv:2208.14474 [hep-ph]}
  \BibitemShut {NoStop}%
\bibitem [{\citenamefont {von Krosigk}\ \emph {et~al.}(2022)\citenamefont {von
  Krosigk} \emph {et~al.}}]{vonKrosigk:2022vnf}%
  \BibitemOpen
  \bibfield  {author} {\bibinfo {author} {\bibfnamefont {B.}~\bibnamefont {von
  Krosigk}} \emph {et~al.},\ }in\ \href@noop {} {\emph {\bibinfo {booktitle}
  {{14th International Workshop on the Identification of Dark Matter 2022}}}}\
  (\bibinfo {year} {2022})\ \Eprint {https://arxiv.org/abs/2209.10950}
  {arXiv:2209.10950 [hep-ex]} \BibitemShut {NoStop}%
\bibitem [{\citenamefont {Seidel}\ and\ \citenamefont
  {Enss}(2022)}]{Seidel:2022ofd}%
  \BibitemOpen
  \bibfield  {author} {\bibinfo {author} {\bibfnamefont {G.~M.}\ \bibnamefont
  {Seidel}}\ and\ \bibinfo {author} {\bibfnamefont {C.}~\bibnamefont {Enss}},\
  }\href@noop {} {\  (\bibinfo {year} {2022})},\ \Eprint
  {https://arxiv.org/abs/2210.06283} {arXiv:2210.06283 [astro-ph.IM]}
  \BibitemShut {NoStop}%
\bibitem [{\citenamefont {Knapen}\ \emph {et~al.}(2018)\citenamefont {Knapen},
  \citenamefont {Lin}, \citenamefont {Pyle},\ and\ \citenamefont
  {Zurek}}]{Knapen:2017ekk}%
  \BibitemOpen
  \bibfield  {author} {\bibinfo {author} {\bibfnamefont {S.}~\bibnamefont
  {Knapen}}, \bibinfo {author} {\bibfnamefont {T.}~\bibnamefont {Lin}},
  \bibinfo {author} {\bibfnamefont {M.}~\bibnamefont {Pyle}},\ and\ \bibinfo
  {author} {\bibfnamefont {K.~M.}\ \bibnamefont {Zurek}},\ }\href
  {https://doi.org/10.1016/j.physletb.2018.08.064} {\bibfield  {journal}
  {\bibinfo  {journal} {Phys. Lett. B}\ }\textbf {\bibinfo {volume} {785}},\
  \bibinfo {pages} {386} (\bibinfo {year} {2018})},\ \Eprint
  {https://arxiv.org/abs/1712.06598} {arXiv:1712.06598 [hep-ph]} \BibitemShut
  {NoStop}%
\bibitem [{\citenamefont {Griffin}\ \emph {et~al.}(2018)\citenamefont
  {Griffin}, \citenamefont {Knapen}, \citenamefont {Lin},\ and\ \citenamefont
  {Zurek}}]{Griffin:2018bjn}%
  \BibitemOpen
  \bibfield  {author} {\bibinfo {author} {\bibfnamefont {S.}~\bibnamefont
  {Griffin}}, \bibinfo {author} {\bibfnamefont {S.}~\bibnamefont {Knapen}},
  \bibinfo {author} {\bibfnamefont {T.}~\bibnamefont {Lin}},\ and\ \bibinfo
  {author} {\bibfnamefont {K.~M.}\ \bibnamefont {Zurek}},\ }\href
  {https://doi.org/10.1103/PhysRevD.98.115034} {\bibfield  {journal} {\bibinfo
  {journal} {Phys. Rev. D}\ }\textbf {\bibinfo {volume} {98}},\ \bibinfo
  {pages} {115034} (\bibinfo {year} {2018})},\ \Eprint
  {https://arxiv.org/abs/1807.10291} {arXiv:1807.10291 [hep-ph]} \BibitemShut
  {NoStop}%
\bibitem [{\citenamefont {Campbell-Deem}\ \emph {et~al.}(2020)\citenamefont
  {Campbell-Deem}, \citenamefont {Cox}, \citenamefont {Knapen}, \citenamefont
  {Lin},\ and\ \citenamefont {Melia}}]{Campbell-Deem:2019hdx}%
  \BibitemOpen
  \bibfield  {author} {\bibinfo {author} {\bibfnamefont {B.}~\bibnamefont
  {Campbell-Deem}}, \bibinfo {author} {\bibfnamefont {P.}~\bibnamefont {Cox}},
  \bibinfo {author} {\bibfnamefont {S.}~\bibnamefont {Knapen}}, \bibinfo
  {author} {\bibfnamefont {T.}~\bibnamefont {Lin}},\ and\ \bibinfo {author}
  {\bibfnamefont {T.}~\bibnamefont {Melia}},\ }\href
  {https://doi.org/10.1103/PhysRevD.101.036006} {\bibfield  {journal} {\bibinfo
   {journal} {Phys. Rev. D}\ }\textbf {\bibinfo {volume} {101}},\ \bibinfo
  {pages} {036006} (\bibinfo {year} {2020})},\ \bibinfo {note} {[Erratum:
  Phys.Rev.D 102, 019904 (2020)]},\ \Eprint {https://arxiv.org/abs/1911.03482}
  {arXiv:1911.03482 [hep-ph]} \BibitemShut {NoStop}%
\bibitem [{\citenamefont {Cox}\ \emph {et~al.}(2019)\citenamefont {Cox},
  \citenamefont {Melia},\ and\ \citenamefont {Rajendran}}]{Cox:2019cod}%
  \BibitemOpen
  \bibfield  {author} {\bibinfo {author} {\bibfnamefont {P.}~\bibnamefont
  {Cox}}, \bibinfo {author} {\bibfnamefont {T.}~\bibnamefont {Melia}},\ and\
  \bibinfo {author} {\bibfnamefont {S.}~\bibnamefont {Rajendran}},\ }\href
  {https://doi.org/10.1103/PhysRevD.100.055011} {\bibfield  {journal} {\bibinfo
   {journal} {Phys. Rev. D}\ }\textbf {\bibinfo {volume} {100}},\ \bibinfo
  {pages} {055011} (\bibinfo {year} {2019})},\ \Eprint
  {https://arxiv.org/abs/1905.05575} {arXiv:1905.05575 [hep-ph]} \BibitemShut
  {NoStop}%
\bibitem [{\citenamefont {Campbell-Deem}\ \emph {et~al.}(2022)\citenamefont
  {Campbell-Deem}, \citenamefont {Knapen}, \citenamefont {Lin},\ and\
  \citenamefont {Villarama}}]{Campbell-Deem:2022fqm}%
  \BibitemOpen
  \bibfield  {author} {\bibinfo {author} {\bibfnamefont {B.}~\bibnamefont
  {Campbell-Deem}}, \bibinfo {author} {\bibfnamefont {S.}~\bibnamefont
  {Knapen}}, \bibinfo {author} {\bibfnamefont {T.}~\bibnamefont {Lin}},\ and\
  \bibinfo {author} {\bibfnamefont {E.}~\bibnamefont {Villarama}},\ }\href
  {https://doi.org/10.1103/PhysRevD.106.036019} {\bibfield  {journal} {\bibinfo
   {journal} {Phys. Rev. D}\ }\textbf {\bibinfo {volume} {106}},\ \bibinfo
  {pages} {036019} (\bibinfo {year} {2022})},\ \Eprint
  {https://arxiv.org/abs/2205.02250} {arXiv:2205.02250 [hep-ph]} \BibitemShut
  {NoStop}%
\bibitem [{\citenamefont {Griffin}\ \emph {et~al.}(2020)\citenamefont
  {Griffin}, \citenamefont {Inzani}, \citenamefont {Trickle}, \citenamefont
  {Zhang},\ and\ \citenamefont {Zurek}}]{Griffin:2019mvc}%
  \BibitemOpen
  \bibfield  {author} {\bibinfo {author} {\bibfnamefont {S.~M.}\ \bibnamefont
  {Griffin}}, \bibinfo {author} {\bibfnamefont {K.}~\bibnamefont {Inzani}},
  \bibinfo {author} {\bibfnamefont {T.}~\bibnamefont {Trickle}}, \bibinfo
  {author} {\bibfnamefont {Z.}~\bibnamefont {Zhang}},\ and\ \bibinfo {author}
  {\bibfnamefont {K.~M.}\ \bibnamefont {Zurek}},\ }\href
  {https://doi.org/10.1103/PhysRevD.101.055004} {\bibfield  {journal} {\bibinfo
   {journal} {Phys. Rev. D}\ }\textbf {\bibinfo {volume} {101}},\ \bibinfo
  {pages} {055004} (\bibinfo {year} {2020})},\ \Eprint
  {https://arxiv.org/abs/1910.10716} {arXiv:1910.10716 [hep-ph]} \BibitemShut
  {NoStop}%
\bibitem [{\citenamefont {Trickle}\ \emph
  {et~al.}(2020{\natexlab{a}})\citenamefont {Trickle}, \citenamefont {Zhang},
  \citenamefont {Zurek}, \citenamefont {Inzani},\ and\ \citenamefont
  {Griffin}}]{Trickle:2019nya}%
  \BibitemOpen
  \bibfield  {author} {\bibinfo {author} {\bibfnamefont {T.}~\bibnamefont
  {Trickle}}, \bibinfo {author} {\bibfnamefont {Z.}~\bibnamefont {Zhang}},
  \bibinfo {author} {\bibfnamefont {K.~M.}\ \bibnamefont {Zurek}}, \bibinfo
  {author} {\bibfnamefont {K.}~\bibnamefont {Inzani}},\ and\ \bibinfo {author}
  {\bibfnamefont {S.~M.}\ \bibnamefont {Griffin}},\ }\href
  {https://doi.org/10.1007/JHEP03(2020)036} {\bibfield  {journal} {\bibinfo
  {journal} {JHEP}\ }\textbf {\bibinfo {volume} {03}},\ \bibinfo {pages}
  {036}},\ \Eprint {https://arxiv.org/abs/1910.08092} {arXiv:1910.08092
  [hep-ph]} \BibitemShut {NoStop}%
\bibitem [{\citenamefont {Kahn}\ and\ \citenamefont
  {Lin}(2022)}]{Kahn:2021ttr}%
  \BibitemOpen
  \bibfield  {author} {\bibinfo {author} {\bibfnamefont {Y.}~\bibnamefont
  {Kahn}}\ and\ \bibinfo {author} {\bibfnamefont {T.}~\bibnamefont {Lin}},\
  }\href {https://doi.org/10.1088/1361-6633/ac5f63} {\bibfield  {journal}
  {\bibinfo  {journal} {Rept. Prog. Phys.}\ }\textbf {\bibinfo {volume} {85}},\
  \bibinfo {pages} {066901} (\bibinfo {year} {2022})},\ \Eprint
  {https://arxiv.org/abs/2108.03239} {arXiv:2108.03239 [hep-ph]} \BibitemShut
  {NoStop}%
\bibitem [{\citenamefont {Maris}\ \emph {et~al.}(2017)\citenamefont {Maris},
  \citenamefont {Seidel},\ and\ \citenamefont {Stein}}]{Maris:2017xvi}%
  \BibitemOpen
  \bibfield  {author} {\bibinfo {author} {\bibfnamefont {H.~J.}\ \bibnamefont
  {Maris}}, \bibinfo {author} {\bibfnamefont {G.~M.}\ \bibnamefont {Seidel}},\
  and\ \bibinfo {author} {\bibfnamefont {D.}~\bibnamefont {Stein}},\ }\href
  {https://doi.org/10.1103/PhysRevLett.119.181303} {\bibfield  {journal}
  {\bibinfo  {journal} {Phys. Rev. Lett.}\ }\textbf {\bibinfo {volume} {119}},\
  \bibinfo {pages} {181303} (\bibinfo {year} {2017})},\ \Eprint
  {https://arxiv.org/abs/1706.00117} {arXiv:1706.00117 [astro-ph.IM]}
  \BibitemShut {NoStop}%
\bibitem [{\citenamefont {Osterman}\ \emph {et~al.}(2020)\citenamefont
  {Osterman}, \citenamefont {Maris}, \citenamefont {Seidel},\ and\
  \citenamefont {Stein}}]{Osterman:2020xkb}%
  \BibitemOpen
  \bibfield  {author} {\bibinfo {author} {\bibfnamefont {D.}~\bibnamefont
  {Osterman}}, \bibinfo {author} {\bibfnamefont {H.}~\bibnamefont {Maris}},
  \bibinfo {author} {\bibfnamefont {G.}~\bibnamefont {Seidel}},\ and\ \bibinfo
  {author} {\bibfnamefont {D.}~\bibnamefont {Stein}},\ }\href
  {https://doi.org/10.1088/1742-6596/1468/1/012071} {\bibfield  {journal}
  {\bibinfo  {journal} {J. Phys. Conf. Ser.}\ }\textbf {\bibinfo {volume}
  {1468}},\ \bibinfo {pages} {012071} (\bibinfo {year} {2020})}\BibitemShut
  {NoStop}%
\bibitem [{\citenamefont {Lyon}\ \emph {et~al.}(2022)\citenamefont {Lyon},
  \citenamefont {Castoria}, \citenamefont {Kleinbaum}, \citenamefont {Qin},
  \citenamefont {Persaud}, \citenamefont {Schenkel},\ and\ \citenamefont
  {Zurek}}]{Lyon:2022sza}%
  \BibitemOpen
  \bibfield  {author} {\bibinfo {author} {\bibfnamefont {S.~A.}\ \bibnamefont
  {Lyon}}, \bibinfo {author} {\bibfnamefont {K.}~\bibnamefont {Castoria}},
  \bibinfo {author} {\bibfnamefont {E.}~\bibnamefont {Kleinbaum}}, \bibinfo
  {author} {\bibfnamefont {Z.}~\bibnamefont {Qin}}, \bibinfo {author}
  {\bibfnamefont {A.}~\bibnamefont {Persaud}}, \bibinfo {author} {\bibfnamefont
  {T.}~\bibnamefont {Schenkel}},\ and\ \bibinfo {author} {\bibfnamefont
  {K.}~\bibnamefont {Zurek}},\ }\href@noop {} {\  (\bibinfo {year} {2022})},\
  \Eprint {https://arxiv.org/abs/2201.00738} {arXiv:2201.00738 [hep-ex]}
  \BibitemShut {NoStop}%
\bibitem [{\citenamefont {Das}\ \emph {et~al.}(2022)\citenamefont {Das},
  \citenamefont {Kurinsky},\ and\ \citenamefont {Leane}}]{Das:2022srn}%
  \BibitemOpen
  \bibfield  {author} {\bibinfo {author} {\bibfnamefont {A.}~\bibnamefont
  {Das}}, \bibinfo {author} {\bibfnamefont {N.}~\bibnamefont {Kurinsky}},\ and\
  \bibinfo {author} {\bibfnamefont {R.~K.}\ \bibnamefont {Leane}},\ }\href@noop
  {} {\  (\bibinfo {year} {2022})},\ \Eprint {https://arxiv.org/abs/2210.09313}
  {arXiv:2210.09313 [hep-ph]} \BibitemShut {NoStop}%
\bibitem [{\citenamefont {Trickle}\ \emph
  {et~al.}(2020{\natexlab{b}})\citenamefont {Trickle}, \citenamefont {Zhang},\
  and\ \citenamefont {Zurek}}]{Trickle:2019ovy}%
  \BibitemOpen
  \bibfield  {author} {\bibinfo {author} {\bibfnamefont {T.}~\bibnamefont
  {Trickle}}, \bibinfo {author} {\bibfnamefont {Z.}~\bibnamefont {Zhang}},\
  and\ \bibinfo {author} {\bibfnamefont {K.~M.}\ \bibnamefont {Zurek}},\ }\href
  {https://doi.org/10.1103/PhysRevLett.124.201801} {\bibfield  {journal}
  {\bibinfo  {journal} {Phys. Rev. Lett.}\ }\textbf {\bibinfo {volume} {124}},\
  \bibinfo {pages} {201801} (\bibinfo {year} {2020}{\natexlab{b}})},\ \Eprint
  {https://arxiv.org/abs/1905.13744} {arXiv:1905.13744 [hep-ph]} \BibitemShut
  {NoStop}%
\bibitem [{\citenamefont {Mitridate}\ \emph {et~al.}(2020)\citenamefont
  {Mitridate}, \citenamefont {Trickle}, \citenamefont {Zhang},\ and\
  \citenamefont {Zurek}}]{Mitridate:2020kly}%
  \BibitemOpen
  \bibfield  {author} {\bibinfo {author} {\bibfnamefont {A.}~\bibnamefont
  {Mitridate}}, \bibinfo {author} {\bibfnamefont {T.}~\bibnamefont {Trickle}},
  \bibinfo {author} {\bibfnamefont {Z.}~\bibnamefont {Zhang}},\ and\ \bibinfo
  {author} {\bibfnamefont {K.~M.}\ \bibnamefont {Zurek}},\ }\href
  {https://doi.org/10.1103/PhysRevD.102.095005} {\bibfield  {journal} {\bibinfo
   {journal} {Phys. Rev. D}\ }\textbf {\bibinfo {volume} {102}},\ \bibinfo
  {pages} {095005} (\bibinfo {year} {2020})},\ \Eprint
  {https://arxiv.org/abs/2005.10256} {arXiv:2005.10256 [hep-ph]} \BibitemShut
  {NoStop}%
\bibitem [{\citenamefont {Chigusa}\ \emph {et~al.}(2020)\citenamefont
  {Chigusa}, \citenamefont {Moroi},\ and\ \citenamefont
  {Nakayama}}]{Chigusa:2020gfs}%
  \BibitemOpen
  \bibfield  {author} {\bibinfo {author} {\bibfnamefont {S.}~\bibnamefont
  {Chigusa}}, \bibinfo {author} {\bibfnamefont {T.}~\bibnamefont {Moroi}},\
  and\ \bibinfo {author} {\bibfnamefont {K.}~\bibnamefont {Nakayama}},\ }\href
  {https://doi.org/10.1103/PhysRevD.101.096013} {\bibfield  {journal} {\bibinfo
   {journal} {Phys. Rev. D}\ }\textbf {\bibinfo {volume} {101}},\ \bibinfo
  {pages} {096013} (\bibinfo {year} {2020})},\ \Eprint
  {https://arxiv.org/abs/2001.10666} {arXiv:2001.10666 [hep-ph]} \BibitemShut
  {NoStop}%
\bibitem [{\citenamefont {Pavaskar}\ \emph {et~al.}(2022)\citenamefont
  {Pavaskar}, \citenamefont {Penco},\ and\ \citenamefont
  {Rothstein}}]{Pavaskar:2021pfo}%
  \BibitemOpen
  \bibfield  {author} {\bibinfo {author} {\bibfnamefont {S.}~\bibnamefont
  {Pavaskar}}, \bibinfo {author} {\bibfnamefont {R.}~\bibnamefont {Penco}},\
  and\ \bibinfo {author} {\bibfnamefont {I.~Z.}\ \bibnamefont {Rothstein}},\
  }\href {https://doi.org/10.21468/SciPostPhys.12.5.155} {\bibfield  {journal}
  {\bibinfo  {journal} {SciPost Phys.}\ }\textbf {\bibinfo {volume} {12}},\
  \bibinfo {pages} {155} (\bibinfo {year} {2022})},\ \Eprint
  {https://arxiv.org/abs/2112.13873} {arXiv:2112.13873 [hep-th]} \BibitemShut
  {NoStop}%
\bibitem [{\citenamefont {Lachance-Quirion}\ \emph {et~al.}(2020)\citenamefont
  {Lachance-Quirion}, \citenamefont {Wolski}, \citenamefont {Tabuchi},
  \citenamefont {Kono}, \citenamefont {Usami},\ and\ \citenamefont
  {Nakamura}}]{doi:10.1126/science.aaz9236}%
  \BibitemOpen
  \bibfield  {author} {\bibinfo {author} {\bibfnamefont {D.}~\bibnamefont
  {Lachance-Quirion}}, \bibinfo {author} {\bibfnamefont {S.~P.}\ \bibnamefont
  {Wolski}}, \bibinfo {author} {\bibfnamefont {Y.}~\bibnamefont {Tabuchi}},
  \bibinfo {author} {\bibfnamefont {S.}~\bibnamefont {Kono}}, \bibinfo {author}
  {\bibfnamefont {K.}~\bibnamefont {Usami}},\ and\ \bibinfo {author}
  {\bibfnamefont {Y.}~\bibnamefont {Nakamura}},\ }\href
  {https://doi.org/10.1126/science.aaz9236} {\bibfield  {journal} {\bibinfo
  {journal} {Science}\ }\textbf {\bibinfo {volume} {367}},\ \bibinfo {pages}
  {425} (\bibinfo {year} {2020})}\BibitemShut {NoStop}%
\bibitem [{\citenamefont {Lachance-Quirion}\ \emph {et~al.}(2019)\citenamefont
  {Lachance-Quirion}, \citenamefont {Tabuchi}, \citenamefont {Gloppe},
  \citenamefont {Usami},\ and\ \citenamefont {Nakamura}}]{lachance2019hybrid}%
  \BibitemOpen
  \bibfield  {author} {\bibinfo {author} {\bibfnamefont {D.}~\bibnamefont
  {Lachance-Quirion}}, \bibinfo {author} {\bibfnamefont {Y.}~\bibnamefont
  {Tabuchi}}, \bibinfo {author} {\bibfnamefont {A.}~\bibnamefont {Gloppe}},
  \bibinfo {author} {\bibfnamefont {K.}~\bibnamefont {Usami}},\ and\ \bibinfo
  {author} {\bibfnamefont {Y.}~\bibnamefont {Nakamura}},\ }\href
  {https://iopscience.iop.org/article/10.7567/1882-0786/ab248d/meta} {\bibfield
   {journal} {\bibinfo  {journal} {Applied Physics Express}\ }\textbf {\bibinfo
  {volume} {12}},\ \bibinfo {pages} {070101} (\bibinfo {year}
  {2019})}\BibitemShut {NoStop}%
\bibitem [{\citenamefont {Lachance-Quirion}\ \emph {et~al.}(2017)\citenamefont
  {Lachance-Quirion}, \citenamefont {Tabuchi}, \citenamefont {Ishino},
  \citenamefont {Noguchi}, \citenamefont {Ishikawa}, \citenamefont {Yamazaki},\
  and\ \citenamefont {Nakamura}}]{doi:10.1126/sciadv.1603150}%
  \BibitemOpen
  \bibfield  {author} {\bibinfo {author} {\bibfnamefont {D.}~\bibnamefont
  {Lachance-Quirion}}, \bibinfo {author} {\bibfnamefont {Y.}~\bibnamefont
  {Tabuchi}}, \bibinfo {author} {\bibfnamefont {S.}~\bibnamefont {Ishino}},
  \bibinfo {author} {\bibfnamefont {A.}~\bibnamefont {Noguchi}}, \bibinfo
  {author} {\bibfnamefont {T.}~\bibnamefont {Ishikawa}}, \bibinfo {author}
  {\bibfnamefont {R.}~\bibnamefont {Yamazaki}},\ and\ \bibinfo {author}
  {\bibfnamefont {Y.}~\bibnamefont {Nakamura}},\ }\href
  {https://doi.org/10.1126/sciadv.1603150} {\bibfield  {journal} {\bibinfo
  {journal} {Science Advances}\ }\textbf {\bibinfo {volume} {3}},\ \bibinfo
  {pages} {e1603150} (\bibinfo {year} {2017})}\BibitemShut {NoStop}%
\bibitem [{\citenamefont {Srivastava}\ and\ \citenamefont
  {Aiyar}(1987)}]{srivastava1987spin}%
  \BibitemOpen
  \bibfield  {author} {\bibinfo {author} {\bibfnamefont {C.}~\bibnamefont
  {Srivastava}}\ and\ \bibinfo {author} {\bibfnamefont {R.}~\bibnamefont
  {Aiyar}},\ }\href
  {https://iopscience.iop.org/article/10.1088/0022-3719/20/8/013/meta?casa_token=oSN3_ZlZiSkAAAAA:6gx-AYLV2joZpWoFB88GH-_6ij3v_r1udRtTGWsnTHDJPsGH4W6FvGbaqZa10-66qGCb0Cu975tMD3cR5TU}
  {\bibfield  {journal} {\bibinfo  {journal} {Journal of Physics C: Solid State
  Physics}\ }\textbf {\bibinfo {volume} {20}},\ \bibinfo {pages} {1119}
  (\bibinfo {year} {1987})}\BibitemShut {NoStop}%
\bibitem [{\citenamefont {Pajda}\ \emph {et~al.}(2001)\citenamefont {Pajda},
  \citenamefont {Kudrnovsk\'y}, \citenamefont {Turek}, \citenamefont {Drchal},\
  and\ \citenamefont {Bruno}}]{PhysRevB.64.174402}%
  \BibitemOpen
  \bibfield  {author} {\bibinfo {author} {\bibfnamefont {M.}~\bibnamefont
  {Pajda}}, \bibinfo {author} {\bibfnamefont {J.}~\bibnamefont {Kudrnovsk\'y}},
  \bibinfo {author} {\bibfnamefont {I.}~\bibnamefont {Turek}}, \bibinfo
  {author} {\bibfnamefont {V.}~\bibnamefont {Drchal}},\ and\ \bibinfo {author}
  {\bibfnamefont {P.}~\bibnamefont {Bruno}},\ }\href
  {https://doi.org/10.1103/PhysRevB.64.174402} {\bibfield  {journal} {\bibinfo
  {journal} {Phys. Rev. B}\ }\textbf {\bibinfo {volume} {64}},\ \bibinfo
  {pages} {174402} (\bibinfo {year} {2001})}\BibitemShut {NoStop}%
\bibitem [{\citenamefont {Burgess}(2000)}]{Burgess:1998ku}%
  \BibitemOpen
  \bibfield  {author} {\bibinfo {author} {\bibfnamefont {C.~P.}\ \bibnamefont
  {Burgess}},\ }\href {https://doi.org/10.1016/S0370-1573(99)00111-8}
  {\bibfield  {journal} {\bibinfo  {journal} {Phys. Rept.}\ }\textbf {\bibinfo
  {volume} {330}},\ \bibinfo {pages} {193} (\bibinfo {year} {2000})},\ \Eprint
  {https://arxiv.org/abs/hep-th/9808176} {arXiv:hep-th/9808176} \BibitemShut
  {NoStop}%
\bibitem [{\citenamefont {Sch\"utte-Engel}\ \emph {et~al.}(2021)\citenamefont
  {Sch\"utte-Engel}, \citenamefont {Marsh}, \citenamefont {Millar},
  \citenamefont {Sekine}, \citenamefont {Chadha-Day}, \citenamefont {Hoof},
  \citenamefont {Ali}, \citenamefont {Fong}, \citenamefont {Hardy},\ and\
  \citenamefont {\v{S}mejkal}}]{Schutte-Engel:2021bqm}%
  \BibitemOpen
  \bibfield  {author} {\bibinfo {author} {\bibfnamefont {J.}~\bibnamefont
  {Sch\"utte-Engel}}, \bibinfo {author} {\bibfnamefont {D.~J.~E.}\ \bibnamefont
  {Marsh}}, \bibinfo {author} {\bibfnamefont {A.~J.}\ \bibnamefont {Millar}},
  \bibinfo {author} {\bibfnamefont {A.}~\bibnamefont {Sekine}}, \bibinfo
  {author} {\bibfnamefont {F.}~\bibnamefont {Chadha-Day}}, \bibinfo {author}
  {\bibfnamefont {S.}~\bibnamefont {Hoof}}, \bibinfo {author} {\bibfnamefont
  {M.~N.}\ \bibnamefont {Ali}}, \bibinfo {author} {\bibfnamefont {K.-C.}\
  \bibnamefont {Fong}}, \bibinfo {author} {\bibfnamefont {E.}~\bibnamefont
  {Hardy}},\ and\ \bibinfo {author} {\bibfnamefont {L.}~\bibnamefont
  {\v{S}mejkal}},\ }\href {https://doi.org/10.1088/1475-7516/2021/08/066}
  {\bibfield  {journal} {\bibinfo  {journal} {JCAP}\ }\textbf {\bibinfo
  {volume} {08}},\ \bibinfo {pages} {066}},\ \Eprint
  {https://arxiv.org/abs/2102.05366} {arXiv:2102.05366 [hep-ph]} \BibitemShut
  {NoStop}%
\bibitem [{\citenamefont {Squires}(1996)}]{squires1996introduction}%
  \BibitemOpen
  \bibfield  {author} {\bibinfo {author} {\bibfnamefont {G.~L.}\ \bibnamefont
  {Squires}},\ }\href {https://doi.org/10.1017/CBO9781139107808} {\emph
  {\bibinfo {title} {Introduction to the theory of thermal neutron
  scattering}}}\ (\bibinfo  {publisher} {Courier Corporation},\ \bibinfo {year}
  {1996})\BibitemShut {NoStop}%
\bibitem [{\citenamefont {Lovesey}(1984)}]{lovesey1984theory}%
  \BibitemOpen
  \bibfield  {author} {\bibinfo {author} {\bibfnamefont {S.}~\bibnamefont
  {Lovesey}},\ }\href {https://books.google.it/books?id=G5kfAQAAMAAJ} {\emph
  {\bibinfo {title} {Theory of Neutron Scattering from Condensed Matter:
  Nuclear scattering}}},\ International series of monographs on physics\
  (\bibinfo  {publisher} {Clarendon Press},\ \bibinfo {year}
  {1984})\BibitemShut {NoStop}%
\bibitem [{\citenamefont {Dyson}(1956)}]{dyson1956general}%
  \BibitemOpen
  \bibfield  {author} {\bibinfo {author} {\bibfnamefont {F.~J.}\ \bibnamefont
  {Dyson}},\ }\href {https://doi.org/10.1103/PhysRev.102.1217} {\bibfield
  {journal} {\bibinfo  {journal} {Physical review}\ }\textbf {\bibinfo {volume}
  {102}},\ \bibinfo {pages} {1217} (\bibinfo {year} {1956})}\BibitemShut
  {NoStop}%
\bibitem [{\citenamefont {Ashcroft}\ and\ \citenamefont
  {Mermin}(2011)}]{ashcroft2011solid}%
  \BibitemOpen
  \bibfield  {author} {\bibinfo {author} {\bibfnamefont {N.}~\bibnamefont
  {Ashcroft}}\ and\ \bibinfo {author} {\bibfnamefont {N.}~\bibnamefont
  {Mermin}},\ }\href {https://books.google.it/books?id=x\_s\_YAAACAAJ} {\emph
  {\bibinfo {title} {Solid State Physics}}}\ (\bibinfo  {publisher} {Cengage
  Learning},\ \bibinfo {year} {2011})\BibitemShut {NoStop}%
\bibitem [{\citenamefont {Hutchings}\ and\ \citenamefont
  {Samuelsen}(1971)}]{hutchings_inelastic_1971}%
  \BibitemOpen
  \bibfield  {author} {\bibinfo {author} {\bibfnamefont {M.~T.}\ \bibnamefont
  {Hutchings}}\ and\ \bibinfo {author} {\bibfnamefont {E.~J.}\ \bibnamefont
  {Samuelsen}},\ }\href
  {https://doi.org/https://doi.org/10.1016/0038-1098(71)90451-0} {\bibfield
  {journal} {\bibinfo  {journal} {Solid State Communications}\ }\textbf
  {\bibinfo {volume} {9}},\ \bibinfo {pages} {1011} (\bibinfo {year}
  {1971})}\BibitemShut {NoStop}%
\bibitem [{\citenamefont {Pepy}(1974)}]{PEPY1974433}%
  \BibitemOpen
  \bibfield  {author} {\bibinfo {author} {\bibfnamefont {G.}~\bibnamefont
  {Pepy}},\ }\href
  {https://doi.org/https://doi.org/10.1016/S0022-3697(74)80037-5} {\bibfield
  {journal} {\bibinfo  {journal} {Journal of Physics and Chemistry of Solids}\
  }\textbf {\bibinfo {volume} {35}},\ \bibinfo {pages} {433} (\bibinfo {year}
  {1974})}\BibitemShut {NoStop}%
\bibitem [{\citenamefont {Samuelsen}\ \emph {et~al.}(1969)\citenamefont
  {Samuelsen}, \citenamefont {Hutchings},\ and\ \citenamefont
  {Shirane}}]{SAMUELSEN19691043}%
  \BibitemOpen
  \bibfield  {author} {\bibinfo {author} {\bibfnamefont {E.}~\bibnamefont
  {Samuelsen}}, \bibinfo {author} {\bibfnamefont {M.}~\bibnamefont
  {Hutchings}},\ and\ \bibinfo {author} {\bibfnamefont {G.}~\bibnamefont
  {Shirane}},\ }\href
  {https://doi.org/https://doi.org/10.1016/0038-1098(69)90466-9} {\bibfield
  {journal} {\bibinfo  {journal} {Solid State Communications}\ }\textbf
  {\bibinfo {volume} {7}},\ \bibinfo {pages} {1043} (\bibinfo {year}
  {1969})}\BibitemShut {NoStop}%
\bibitem [{\citenamefont {Jain}\ \emph {et~al.}(2013)\citenamefont {Jain},
  \citenamefont {Ong}, \citenamefont {Hautier}, \citenamefont {Chen},
  \citenamefont {Richards}, \citenamefont {Dacek}, \citenamefont {Cholia},
  \citenamefont {Gunter}, \citenamefont {Skinner}, \citenamefont {Ceder},\ and\
  \citenamefont {Persson}}]{doi:10.1063/1.4812323}%
  \BibitemOpen
  \bibfield  {author} {\bibinfo {author} {\bibfnamefont {A.}~\bibnamefont
  {Jain}}, \bibinfo {author} {\bibfnamefont {S.~P.}\ \bibnamefont {Ong}},
  \bibinfo {author} {\bibfnamefont {G.}~\bibnamefont {Hautier}}, \bibinfo
  {author} {\bibfnamefont {W.}~\bibnamefont {Chen}}, \bibinfo {author}
  {\bibfnamefont {W.~D.}\ \bibnamefont {Richards}}, \bibinfo {author}
  {\bibfnamefont {S.}~\bibnamefont {Dacek}}, \bibinfo {author} {\bibfnamefont
  {S.}~\bibnamefont {Cholia}}, \bibinfo {author} {\bibfnamefont
  {D.}~\bibnamefont {Gunter}}, \bibinfo {author} {\bibfnamefont
  {D.}~\bibnamefont {Skinner}}, \bibinfo {author} {\bibfnamefont
  {G.}~\bibnamefont {Ceder}},\ and\ \bibinfo {author} {\bibfnamefont {K.~A.}\
  \bibnamefont {Persson}},\ }\href {https://doi.org/10.1063/1.4812323}
  {\bibfield  {journal} {\bibinfo  {journal} {APL Materials}\ }\textbf
  {\bibinfo {volume} {1}},\ \bibinfo {pages} {011002} (\bibinfo {year}
  {2013})}\BibitemShut {NoStop}%
\bibitem [{\citenamefont {Trickle}\ \emph {et~al.}(2022)\citenamefont
  {Trickle}, \citenamefont {Zhang},\ and\ \citenamefont
  {Zurek}}]{Trickle:2020oki}%
  \BibitemOpen
  \bibfield  {author} {\bibinfo {author} {\bibfnamefont {T.}~\bibnamefont
  {Trickle}}, \bibinfo {author} {\bibfnamefont {Z.}~\bibnamefont {Zhang}},\
  and\ \bibinfo {author} {\bibfnamefont {K.~M.}\ \bibnamefont {Zurek}},\ }\href
  {https://doi.org/10.1103/PhysRevD.105.015001} {\bibfield  {journal} {\bibinfo
   {journal} {Phys. Rev. D}\ }\textbf {\bibinfo {volume} {105}},\ \bibinfo
  {pages} {015001} (\bibinfo {year} {2022})},\ \Eprint
  {https://arxiv.org/abs/2009.13534} {arXiv:2009.13534 [hep-ph]} \BibitemShut
  {NoStop}%
\bibitem [{\citenamefont {Sigurdson}\ \emph {et~al.}(2004)\citenamefont
  {Sigurdson}, \citenamefont {Doran}, \citenamefont {Kurylov}, \citenamefont
  {Caldwell},\ and\ \citenamefont {Kamionkowski}}]{Sigurdson:2004zp}%
  \BibitemOpen
  \bibfield  {author} {\bibinfo {author} {\bibfnamefont {K.}~\bibnamefont
  {Sigurdson}}, \bibinfo {author} {\bibfnamefont {M.}~\bibnamefont {Doran}},
  \bibinfo {author} {\bibfnamefont {A.}~\bibnamefont {Kurylov}}, \bibinfo
  {author} {\bibfnamefont {R.~R.}\ \bibnamefont {Caldwell}},\ and\ \bibinfo
  {author} {\bibfnamefont {M.}~\bibnamefont {Kamionkowski}},\ }\href
  {https://doi.org/10.1103/PhysRevD.70.083501} {\bibfield  {journal} {\bibinfo
  {journal} {Phys. Rev. D}\ }\textbf {\bibinfo {volume} {70}},\ \bibinfo
  {pages} {083501} (\bibinfo {year} {2004})},\ \bibinfo {note} {[Erratum:
  Phys.Rev.D 73, 089903 (2006)]},\ \Eprint
  {https://arxiv.org/abs/astro-ph/0406355} {arXiv:astro-ph/0406355}
  \BibitemShut {NoStop}%
\bibitem [{\citenamefont {Masso}\ \emph {et~al.}(2009)\citenamefont {Masso},
  \citenamefont {Mohanty},\ and\ \citenamefont {Rao}}]{Masso:2009mu}%
  \BibitemOpen
  \bibfield  {author} {\bibinfo {author} {\bibfnamefont {E.}~\bibnamefont
  {Masso}}, \bibinfo {author} {\bibfnamefont {S.}~\bibnamefont {Mohanty}},\
  and\ \bibinfo {author} {\bibfnamefont {S.}~\bibnamefont {Rao}},\ }\href
  {https://doi.org/10.1103/PhysRevD.80.036009} {\bibfield  {journal} {\bibinfo
  {journal} {Phys. Rev. D}\ }\textbf {\bibinfo {volume} {80}},\ \bibinfo
  {pages} {036009} (\bibinfo {year} {2009})},\ \Eprint
  {https://arxiv.org/abs/0906.1979} {arXiv:0906.1979 [hep-ph]} \BibitemShut
  {NoStop}%
\bibitem [{\citenamefont {Chang}\ \emph
  {et~al.}(2010{\natexlab{a}})\citenamefont {Chang}, \citenamefont {Weiner},\
  and\ \citenamefont {Yavin}}]{Chang:2010en}%
  \BibitemOpen
  \bibfield  {author} {\bibinfo {author} {\bibfnamefont {S.}~\bibnamefont
  {Chang}}, \bibinfo {author} {\bibfnamefont {N.}~\bibnamefont {Weiner}},\ and\
  \bibinfo {author} {\bibfnamefont {I.}~\bibnamefont {Yavin}},\ }\href
  {https://doi.org/10.1103/PhysRevD.82.125011} {\bibfield  {journal} {\bibinfo
  {journal} {Phys. Rev. D}\ }\textbf {\bibinfo {volume} {82}},\ \bibinfo
  {pages} {125011} (\bibinfo {year} {2010}{\natexlab{a}})},\ \Eprint
  {https://arxiv.org/abs/1007.4200} {arXiv:1007.4200 [hep-ph]} \BibitemShut
  {NoStop}%
\bibitem [{\citenamefont {Barger}\ \emph {et~al.}(2011)\citenamefont {Barger},
  \citenamefont {Keung},\ and\ \citenamefont {Marfatia}}]{Barger:2010gv}%
  \BibitemOpen
  \bibfield  {author} {\bibinfo {author} {\bibfnamefont {V.}~\bibnamefont
  {Barger}}, \bibinfo {author} {\bibfnamefont {W.-Y.}\ \bibnamefont {Keung}},\
  and\ \bibinfo {author} {\bibfnamefont {D.}~\bibnamefont {Marfatia}},\ }\href
  {https://doi.org/10.1016/j.physletb.2010.12.008} {\bibfield  {journal}
  {\bibinfo  {journal} {Phys. Lett. B}\ }\textbf {\bibinfo {volume} {696}},\
  \bibinfo {pages} {74} (\bibinfo {year} {2011})},\ \Eprint
  {https://arxiv.org/abs/1007.4345} {arXiv:1007.4345 [hep-ph]} \BibitemShut
  {NoStop}%
\bibitem [{\citenamefont {Fitzpatrick}\ \emph {et~al.}(2013)\citenamefont
  {Fitzpatrick}, \citenamefont {Haxton}, \citenamefont {Katz}, \citenamefont
  {Lubbers},\ and\ \citenamefont {Xu}}]{Fitzpatrick:2012ix}%
  \BibitemOpen
  \bibfield  {author} {\bibinfo {author} {\bibfnamefont {A.~L.}\ \bibnamefont
  {Fitzpatrick}}, \bibinfo {author} {\bibfnamefont {W.}~\bibnamefont {Haxton}},
  \bibinfo {author} {\bibfnamefont {E.}~\bibnamefont {Katz}}, \bibinfo {author}
  {\bibfnamefont {N.}~\bibnamefont {Lubbers}},\ and\ \bibinfo {author}
  {\bibfnamefont {Y.}~\bibnamefont {Xu}},\ }\href
  {https://doi.org/10.1088/1475-7516/2013/02/004} {\bibfield  {journal}
  {\bibinfo  {journal} {JCAP}\ }\textbf {\bibinfo {volume} {02}},\ \bibinfo
  {pages} {004}},\ \Eprint {https://arxiv.org/abs/1203.3542} {arXiv:1203.3542
  [hep-ph]} \BibitemShut {NoStop}%
\bibitem [{\citenamefont {Gresham}\ and\ \citenamefont
  {Zurek}(2014)}]{Gresham:2014vja}%
  \BibitemOpen
  \bibfield  {author} {\bibinfo {author} {\bibfnamefont {M.~I.}\ \bibnamefont
  {Gresham}}\ and\ \bibinfo {author} {\bibfnamefont {K.~M.}\ \bibnamefont
  {Zurek}},\ }\href {https://doi.org/10.1103/PhysRevD.89.123521} {\bibfield
  {journal} {\bibinfo  {journal} {Phys. Rev. D}\ }\textbf {\bibinfo {volume}
  {89}},\ \bibinfo {pages} {123521} (\bibinfo {year} {2014})},\ \Eprint
  {https://arxiv.org/abs/1401.3739} {arXiv:1401.3739 [hep-ph]} \BibitemShut
  {NoStop}%
\bibitem [{\citenamefont {Del~Nobile}\ \emph {et~al.}(2014)\citenamefont
  {Del~Nobile}, \citenamefont {Gelmini}, \citenamefont {Gondolo},\ and\
  \citenamefont {Huh}}]{DelNobile:2014eta}%
  \BibitemOpen
  \bibfield  {author} {\bibinfo {author} {\bibfnamefont {E.}~\bibnamefont
  {Del~Nobile}}, \bibinfo {author} {\bibfnamefont {G.~B.}\ \bibnamefont
  {Gelmini}}, \bibinfo {author} {\bibfnamefont {P.}~\bibnamefont {Gondolo}},\
  and\ \bibinfo {author} {\bibfnamefont {J.-H.}\ \bibnamefont {Huh}},\ }\href
  {https://doi.org/10.1088/1475-7516/2014/06/002} {\bibfield  {journal}
  {\bibinfo  {journal} {JCAP}\ }\textbf {\bibinfo {volume} {06}},\ \bibinfo
  {pages} {002}},\ \Eprint {https://arxiv.org/abs/1401.4508} {arXiv:1401.4508
  [hep-ph]} \BibitemShut {NoStop}%
\bibitem [{\citenamefont {Kavanagh}\ \emph {et~al.}(2019)\citenamefont
  {Kavanagh}, \citenamefont {Panci},\ and\ \citenamefont
  {Ziegler}}]{Kavanagh:2018xeh}%
  \BibitemOpen
  \bibfield  {author} {\bibinfo {author} {\bibfnamefont {B.~J.}\ \bibnamefont
  {Kavanagh}}, \bibinfo {author} {\bibfnamefont {P.}~\bibnamefont {Panci}},\
  and\ \bibinfo {author} {\bibfnamefont {R.}~\bibnamefont {Ziegler}},\ }\href
  {https://doi.org/10.1007/JHEP04(2019)089} {\bibfield  {journal} {\bibinfo
  {journal} {JHEP}\ }\textbf {\bibinfo {volume} {04}},\ \bibinfo {pages}
  {089}},\ \Eprint {https://arxiv.org/abs/1810.00033} {arXiv:1810.00033
  [hep-ph]} \BibitemShut {NoStop}%
\bibitem [{\citenamefont {Chu}\ \emph {et~al.}(2019)\citenamefont {Chu},
  \citenamefont {Pradler},\ and\ \citenamefont {Semmelrock}}]{Chu:2018qrm}%
  \BibitemOpen
  \bibfield  {author} {\bibinfo {author} {\bibfnamefont {X.}~\bibnamefont
  {Chu}}, \bibinfo {author} {\bibfnamefont {J.}~\bibnamefont {Pradler}},\ and\
  \bibinfo {author} {\bibfnamefont {L.}~\bibnamefont {Semmelrock}},\ }\href
  {https://doi.org/10.1103/PhysRevD.99.015040} {\bibfield  {journal} {\bibinfo
  {journal} {Phys. Rev. D}\ }\textbf {\bibinfo {volume} {99}},\ \bibinfo
  {pages} {015040} (\bibinfo {year} {2019})},\ \Eprint
  {https://arxiv.org/abs/1811.04095} {arXiv:1811.04095 [hep-ph]} \BibitemShut
  {NoStop}%
\bibitem [{\citenamefont {Banks}\ \emph {et~al.}(2010)\citenamefont {Banks},
  \citenamefont {Fortin},\ and\ \citenamefont {Thomas}}]{Banks:2010eh}%
  \BibitemOpen
  \bibfield  {author} {\bibinfo {author} {\bibfnamefont {T.}~\bibnamefont
  {Banks}}, \bibinfo {author} {\bibfnamefont {J.-F.}\ \bibnamefont {Fortin}},\
  and\ \bibinfo {author} {\bibfnamefont {S.}~\bibnamefont {Thomas}},\
  }\href@noop {} {\  (\bibinfo {year} {2010})},\ \Eprint
  {https://arxiv.org/abs/1007.5515} {arXiv:1007.5515 [hep-ph]} \BibitemShut
  {NoStop}%
\bibitem [{\citenamefont {Bagnasco}\ \emph {et~al.}(1994)\citenamefont
  {Bagnasco}, \citenamefont {Dine},\ and\ \citenamefont
  {Thomas}}]{Bagnasco:1993st}%
  \BibitemOpen
  \bibfield  {author} {\bibinfo {author} {\bibfnamefont {J.}~\bibnamefont
  {Bagnasco}}, \bibinfo {author} {\bibfnamefont {M.}~\bibnamefont {Dine}},\
  and\ \bibinfo {author} {\bibfnamefont {S.~D.}\ \bibnamefont {Thomas}},\
  }\href {https://doi.org/10.1016/0370-2693(94)90830-3} {\bibfield  {journal}
  {\bibinfo  {journal} {Phys. Lett. B}\ }\textbf {\bibinfo {volume} {320}},\
  \bibinfo {pages} {99} (\bibinfo {year} {1994})},\ \Eprint
  {https://arxiv.org/abs/hep-ph/9310290} {arXiv:hep-ph/9310290} \BibitemShut
  {NoStop}%
\bibitem [{\citenamefont {Chang}\ \emph
  {et~al.}(2010{\natexlab{b}})\citenamefont {Chang}, \citenamefont {Pierce},\
  and\ \citenamefont {Weiner}}]{Chang:2009yt}%
  \BibitemOpen
  \bibfield  {author} {\bibinfo {author} {\bibfnamefont {S.}~\bibnamefont
  {Chang}}, \bibinfo {author} {\bibfnamefont {A.}~\bibnamefont {Pierce}},\ and\
  \bibinfo {author} {\bibfnamefont {N.}~\bibnamefont {Weiner}},\ }\href
  {https://doi.org/10.1088/1475-7516/2010/01/006} {\bibfield  {journal}
  {\bibinfo  {journal} {JCAP}\ }\textbf {\bibinfo {volume} {01}},\ \bibinfo
  {pages} {006}},\ \Eprint {https://arxiv.org/abs/0908.3192} {arXiv:0908.3192
  [hep-ph]} \BibitemShut {NoStop}%
\bibitem [{\citenamefont {Manohar}\ and\ \citenamefont
  {Wise}(2000)}]{Manohar:2000dt}%
  \BibitemOpen
  \bibfield  {author} {\bibinfo {author} {\bibfnamefont {A.~V.}\ \bibnamefont
  {Manohar}}\ and\ \bibinfo {author} {\bibfnamefont {M.~B.}\ \bibnamefont
  {Wise}},\ }\href {https://doi.org/10.1017/CBO9780511529351} {\emph {\bibinfo
  {title} {{Heavy quark physics}}}},\ Vol.~\bibinfo {volume} {10}\ (\bibinfo
  {year} {2000})\BibitemShut {NoStop}%
\bibitem [{\citenamefont {Piffl}\ \emph {et~al.}(2014)\citenamefont {Piffl}
  \emph {et~al.}}]{Piffl:2013mla}%
  \BibitemOpen
  \bibfield  {author} {\bibinfo {author} {\bibfnamefont {T.}~\bibnamefont
  {Piffl}} \emph {et~al.},\ }\href
  {https://doi.org/10.1051/0004-6361/201322531} {\bibfield  {journal} {\bibinfo
   {journal} {Astron. Astrophys.}\ }\textbf {\bibinfo {volume} {562}},\
  \bibinfo {pages} {A91} (\bibinfo {year} {2014})},\ \Eprint
  {https://arxiv.org/abs/1309.4293} {arXiv:1309.4293 [astro-ph.GA]}
  \BibitemShut {NoStop}%
\bibitem [{\citenamefont {Monari}\ \emph {et~al.}(2018)\citenamefont {Monari},
  \citenamefont {Famaey}, \citenamefont {Carrillo}, \citenamefont {Piffl},
  \citenamefont {Steinmetz}, \citenamefont {Wyse}, \citenamefont {Anders},
  \citenamefont {Chiappini},\ and\ \citenamefont
  {Jan\ss{}en}}]{Monari:2018ckf}%
  \BibitemOpen
  \bibfield  {author} {\bibinfo {author} {\bibfnamefont {G.}~\bibnamefont
  {Monari}}, \bibinfo {author} {\bibfnamefont {B.}~\bibnamefont {Famaey}},
  \bibinfo {author} {\bibfnamefont {I.}~\bibnamefont {Carrillo}}, \bibinfo
  {author} {\bibfnamefont {T.}~\bibnamefont {Piffl}}, \bibinfo {author}
  {\bibfnamefont {M.}~\bibnamefont {Steinmetz}}, \bibinfo {author}
  {\bibfnamefont {R.~F.~G.}\ \bibnamefont {Wyse}}, \bibinfo {author}
  {\bibfnamefont {F.}~\bibnamefont {Anders}}, \bibinfo {author} {\bibfnamefont
  {C.}~\bibnamefont {Chiappini}},\ and\ \bibinfo {author} {\bibfnamefont
  {K.}~\bibnamefont {Jan\ss{}en}},\ }\href
  {https://doi.org/10.1051/0004-6361/201833748} {\bibfield  {journal} {\bibinfo
   {journal} {Astron. Astrophys.}\ }\textbf {\bibinfo {volume} {616}},\
  \bibinfo {pages} {L9} (\bibinfo {year} {2018})},\ \Eprint
  {https://arxiv.org/abs/1807.04565} {arXiv:1807.04565 [astro-ph.GA]}
  \BibitemShut {NoStop}%
\bibitem [{\citenamefont {Hochberg}\ \emph
  {et~al.}(2016{\natexlab{c}})\citenamefont {Hochberg}, \citenamefont {Pyle},
  \citenamefont {Zhao},\ and\ \citenamefont {Zurek}}]{Hochberg_2016}%
  \BibitemOpen
  \bibfield  {author} {\bibinfo {author} {\bibfnamefont {Y.}~\bibnamefont
  {Hochberg}}, \bibinfo {author} {\bibfnamefont {M.}~\bibnamefont {Pyle}},
  \bibinfo {author} {\bibfnamefont {Y.}~\bibnamefont {Zhao}},\ and\ \bibinfo
  {author} {\bibfnamefont {K.~M.}\ \bibnamefont {Zurek}},\ }\bibfield
  {journal} {\bibinfo  {journal} {Journal of High Energy Physics}\ }\textbf
  {\bibinfo {volume} {2016}},\ \href {https://doi.org/10.1007/jhep08(2016)057}
  {10.1007/jhep08(2016)057} (\bibinfo {year} {2016}{\natexlab{c}})\BibitemShut
  {NoStop}%
\bibitem [{\citenamefont {Kakhidze}\ and\ \citenamefont
  {Kolokolov}(1991)}]{Kakhidze:1990in}%
  \BibitemOpen
  \bibfield  {author} {\bibinfo {author} {\bibfnamefont {A.~I.}\ \bibnamefont
  {Kakhidze}}\ and\ \bibinfo {author} {\bibfnamefont {I.~V.}\ \bibnamefont
  {Kolokolov}},\ }\href@noop {} {\bibfield  {journal} {\bibinfo  {journal}
  {Sov. Phys. JETP}\ }\textbf {\bibinfo {volume} {72}},\ \bibinfo {pages} {598}
  (\bibinfo {year} {1991})}\BibitemShut {NoStop}%
\bibitem [{\citenamefont {Marsh}\ \emph {et~al.}(2019)\citenamefont {Marsh},
  \citenamefont {Fong}, \citenamefont {Lentz}, \citenamefont {Smejkal},\ and\
  \citenamefont {Ali}}]{Marsh:2018dlj}%
  \BibitemOpen
  \bibfield  {author} {\bibinfo {author} {\bibfnamefont {D.~J.~E.}\
  \bibnamefont {Marsh}}, \bibinfo {author} {\bibfnamefont {K.-C.}\ \bibnamefont
  {Fong}}, \bibinfo {author} {\bibfnamefont {E.~W.}\ \bibnamefont {Lentz}},
  \bibinfo {author} {\bibfnamefont {L.}~\bibnamefont {Smejkal}},\ and\ \bibinfo
  {author} {\bibfnamefont {M.~N.}\ \bibnamefont {Ali}},\ }\href
  {https://doi.org/10.1103/PhysRevLett.123.121601} {\bibfield  {journal}
  {\bibinfo  {journal} {Phys. Rev. Lett.}\ }\textbf {\bibinfo {volume} {123}},\
  \bibinfo {pages} {121601} (\bibinfo {year} {2019})},\ \Eprint
  {https://arxiv.org/abs/1807.08810} {arXiv:1807.08810 [hep-ph]} \BibitemShut
  {NoStop}%
\bibitem [{\citenamefont {Barbieri}\ \emph {et~al.}(2017)\citenamefont
  {Barbieri}, \citenamefont {Braggio}, \citenamefont {Carugno}, \citenamefont
  {Gallo}, \citenamefont {Lombardi}, \citenamefont {Ortolan}, \citenamefont
  {Pengo}, \citenamefont {Ruoso},\ and\ \citenamefont
  {Speake}}]{Barbieri:2016vwg}%
  \BibitemOpen
  \bibfield  {author} {\bibinfo {author} {\bibfnamefont {R.}~\bibnamefont
  {Barbieri}}, \bibinfo {author} {\bibfnamefont {C.}~\bibnamefont {Braggio}},
  \bibinfo {author} {\bibfnamefont {G.}~\bibnamefont {Carugno}}, \bibinfo
  {author} {\bibfnamefont {C.~S.}\ \bibnamefont {Gallo}}, \bibinfo {author}
  {\bibfnamefont {A.}~\bibnamefont {Lombardi}}, \bibinfo {author}
  {\bibfnamefont {A.}~\bibnamefont {Ortolan}}, \bibinfo {author} {\bibfnamefont
  {R.}~\bibnamefont {Pengo}}, \bibinfo {author} {\bibfnamefont
  {G.}~\bibnamefont {Ruoso}},\ and\ \bibinfo {author} {\bibfnamefont {C.~C.}\
  \bibnamefont {Speake}},\ }\href {https://doi.org/10.1016/j.dark.2017.01.003}
  {\bibfield  {journal} {\bibinfo  {journal} {Phys. Dark Univ.}\ }\textbf
  {\bibinfo {volume} {15}},\ \bibinfo {pages} {135} (\bibinfo {year} {2017})},\
  \Eprint {https://arxiv.org/abs/1606.02201} {arXiv:1606.02201 [hep-ph]}
  \BibitemShut {NoStop}%
\bibitem [{\citenamefont {Crescini}\ \emph {et~al.}(2018)\citenamefont
  {Crescini} \emph {et~al.}}]{Crescini:2018qrz}%
  \BibitemOpen
  \bibfield  {author} {\bibinfo {author} {\bibfnamefont {N.}~\bibnamefont
  {Crescini}} \emph {et~al.},\ }\href
  {https://doi.org/10.1140/epjc/s10052-018-6163-8} {\bibfield  {journal}
  {\bibinfo  {journal} {Eur. Phys. J. C}\ }\textbf {\bibinfo {volume} {78}},\
  \bibinfo {pages} {703} (\bibinfo {year} {2018})},\ \bibinfo {note} {[Erratum:
  Eur.Phys.J.C 78, 813 (2018)]},\ \Eprint {https://arxiv.org/abs/1806.00310}
  {arXiv:1806.00310 [hep-ex]} \BibitemShut {NoStop}%
\bibitem [{\citenamefont {Crescini}\ \emph {et~al.}(2020)\citenamefont
  {Crescini} \emph {et~al.}}]{QUAX:2020adt}%
  \BibitemOpen
  \bibfield  {author} {\bibinfo {author} {\bibfnamefont {N.}~\bibnamefont
  {Crescini}} \emph {et~al.} (\bibinfo {collaboration} {QUAX}),\ }\href
  {https://doi.org/10.1103/PhysRevLett.124.171801} {\bibfield  {journal}
  {\bibinfo  {journal} {Phys. Rev. Lett.}\ }\textbf {\bibinfo {volume} {124}},\
  \bibinfo {pages} {171801} (\bibinfo {year} {2020})},\ \Eprint
  {https://arxiv.org/abs/2001.08940} {arXiv:2001.08940 [hep-ex]} \BibitemShut
  {NoStop}%
\bibitem [{\citenamefont {Altland}\ and\ \citenamefont
  {Simons}(2010)}]{altland_simons_2010}%
  \BibitemOpen
  \bibfield  {author} {\bibinfo {author} {\bibfnamefont {A.}~\bibnamefont
  {Altland}}\ and\ \bibinfo {author} {\bibfnamefont {B.~D.}\ \bibnamefont
  {Simons}},\ }\href {https://doi.org/10.1017/CBO9780511789984} {\emph
  {\bibinfo {title} {Condensed Matter Field Theory}}},\ \bibinfo {edition}
  {2nd}\ ed.\ (\bibinfo  {publisher} {Cambridge University Press},\ \bibinfo
  {year} {2010})\BibitemShut {NoStop}%
\bibitem [{\citenamefont {Hutchings}\ and\ \citenamefont
  {Samuelsen}(1972)}]{hutchings1972measurement}%
  \BibitemOpen
  \bibfield  {author} {\bibinfo {author} {\bibfnamefont {M.~T.}\ \bibnamefont
  {Hutchings}}\ and\ \bibinfo {author} {\bibfnamefont {E.}~\bibnamefont
  {Samuelsen}},\ }\href {https://doi.org/10.1103/PhysRevB.6.3447} {\bibfield
  {journal} {\bibinfo  {journal} {Physical Review B}\ }\textbf {\bibinfo
  {volume} {6}},\ \bibinfo {pages} {3447} (\bibinfo {year} {1972})}\BibitemShut
  {NoStop}%
\end{thebibliography}%

%%%%%%%%%%%%%%%%%%%%%%%%%%%%%%%%%%%%%%%%%%%%%%%%%%%

\end{document}